\documentclass{iopjournal}

\pdfoutput=1
\usepackage[utf8]{inputenc}
\usepackage[english]{babel}
\usepackage[T1]{fontenc}
\usepackage{lmodern}

\usepackage{amsmath}
\usepackage{hyperref}
\usepackage{enumitem}
\usepackage{colortbl}
\usepackage{tabularx}
\usepackage{etoolbox}
\usepackage{tcolorbox}
\usepackage{tikz}
\usepackage{lipsum}
\allowdisplaybreaks
\usepackage{mathtools}
\usepackage{amsmath,amsfonts,amssymb,amsthm,calc}
\usepackage{commath}
\usepackage{physics}
\usepackage{caption}
\usepackage{subcaption}

\newtheorem{theorem}{Theorem}
\newtheorem{lemma}{Lemma}

\newtheorem{corollary}{Corollary}
\newtheorem{defn}{Definition}
\newtheorem{protocol}{Protocol}
\newtheorem{conjecture}{Conjecture}

\newtheoremstyle{theorem}{4mm}{1mm}{\itshape}{ }{\bfseries}{.}{ }{}

\newcommand{\Mod}[1]{\ (\mathrm{mod}\ #1)}
\usepackage{thmtools}

\newenvironment{proofof}[1]{\begin{trivlist}\item[]{\flushleft\bf 
Proof of~#1 }}
{\qed\end{trivlist}}

\begin{document}

\articletype{Paper} 

\title{Bounds on Multipartite Nonlocality via Reduction to Biased Nonlocality}

\author{Hafiza Rumlah Amer$^1$\orcid{0009-0002-9910-4098} and Jibran Rashid$^{2}$\orcid{0000-0002-6927-7417}}

\affil{$^{1,2}$School of Mathematics and Computer Science, Institute of Business Administration, Karachi, Pakistan}

\email{hramer@iba.edu.pk, jrashid@iba.edu.pk}


\begin{abstract}
 Multipartite information principles are needed to understand nonlocal quantum correlations. Towards that end, we provide optimal bounds on genuine multipartite nonlocality for classes of $THRESHOLD$ games using the LOCCG (Local Operations and Classical Communication with Grouping) model. Our proof develops a reduction between multipartite nonlocal and biased bipartite nonlocal games. Generalizing this reduction to a larger class of games may build a bridge from multipartite to bipartite principles.
\end{abstract}

\section{Introduction}\label{int}
A three player Bell scenario for binary inputs $x,y,z$ and outputs $a,b,c$ respectively, corresponds to a joint conditional probability distribution $p(abc | xyz)$. The distribution is considered a genuine tripartite nonlocal correlation~\cite{Bancal13} if it is not  possible to produce $p(abc | xyz)$ by \emph{any} two of the three players jointly acting together. Such distributions are characterized with a decomposition given by  
\begin{align}\label{eq:svetl}
    p(abc | xyz ) =  \sum_{\lambda} q_{\lambda} \, 
    p_{\lambda}(a|x) \, p_{\lambda}(bc |yz) 
    + \sum_{\mu} q_{\mu} \, p_{\mu}(b|y) \, p_{\mu}(ac |xz) 
    + \sum_{\nu} q_{\nu} \, p_{\nu}(c|z) \, p_{\nu}(ab |xy),   
\end{align}
where $0 \leq q_{\lambda}, q_{\mu}, q_{\nu} \leq 1$ and $\sum_{\lambda} q_{\lambda} + \sum_{\mu} q_{\mu} + \sum_{\nu} q_{\nu}=1$. The distribution is a convex combination of all possibilities of two players being grouped together.

\begin{figure}
\begin{center}
\begin{subfigure}{.5\textwidth}
   \centering\captionsetup{width=.8\linewidth}
\includegraphics[width=7.5cm, height=3cm]{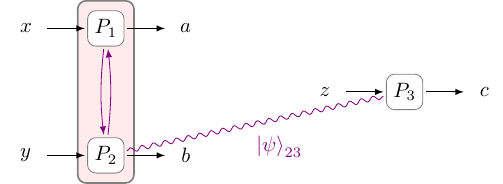}
\caption{Communication is allowed between $P_1$ and $P_2$ while $P_2$ and $P_3$ share entanglement.}
  \label{fig:sub1}
\end{subfigure}%
\begin{subfigure}{.5\textwidth}
\centering\captionsetup{width=.8\linewidth}
\includegraphics[width=7.5cm, height=3cm]{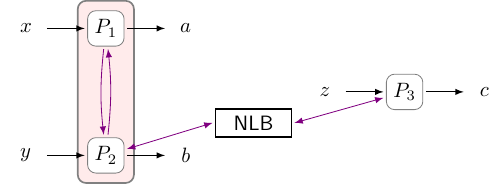}
\caption{Communication is allowed between $P_1$ and $P_2$ while $P_2$ and $P_3$ share an NLB.}
  \label{fig:sub2}
\end{subfigure}
\end{center}
\caption{\label{fig:loccgtri} Three player Bell scenario where players $1$ and $2$ are grouped together via communication while players $2$ and $3$ share a different bipartite resource.}
\label{fig:trisub}
\end{figure}
Consider now whether it is possible to operationally manifest a distribution $p(abc | xyz)$ where any of the two players jointly act together but it cannot be decomposed in the form of Equation~\ref{eq:svetl}. Figure~\ref{fig:loccgtri} shows two situations where players $1$ and $2$ are grouped together via communication and players $2$ and $3$ share either an entangled state or a nonlocal box. Note that even though we have multiple groupings, we still satisfy the requirement that only $2$ out of the $3$ players are grouped together using different resources.  However, the players can now produce a distribution $p(abc | xyz )$ of the form in Equation~\ref{eq:loccgtri}. Note that the expression cannot be decomposed in the form of Equation \ref{eq:svetl} even though it is a distribution that is not genuinely tripartite. 
{\small
\begin{equation}\label{eq:loccgtri}
    p(abc | xyz ) =  \hphantom{+} \sum_{\lambda} q_{\lambda} \, 
    p_{\lambda}(a|xy) \, p_{\lambda}(bc |xyz)     + \sum_{\mu} q_{\mu} \, p_{\mu}(b|yz) \, p_{\mu}(ac |xyz) + \sum_{\nu} q_{\nu} \, p_{\nu}(c|xz) \, p_{\nu}(ab |xyz).   
\end{equation}
}
As an example, consider the three player $AND$ game with winning criterion $ a \oplus b \oplus c =x \wedge y \wedge z $. If~we fix the grouping as shown in Figure~\ref{fig:trisub} and only player $1$ communicates input $x$ to player $2$, then the distribution structure simplifies to
\begin{equation} \label{eq:nonsvet}
    p(abc|xyz) = \sum_{\lambda} q_{\lambda} \, p_{\lambda}(a|x) \, p_{\lambda}(bc |xyz),
\end{equation}
where $\sum_{\lambda} q_{\lambda}=1$. Note that Equation~\ref{eq:nonsvet} does not satisfy the form of Equation~\ref{eq:svetl}, although it cannot be classified as genuinely tripartite. \emph{We can conclude this by observing that player $1$ is local in the distribution}. The players in Figure~\ref{fig:sub2} can win the game with certainty if player $1$ outputs $0$ and players $2$ and $3$ provide the output of the perfect nonlocal box with inputs $x \wedge y$ and $z$, respectively. For the quantum strategy in Figure~\ref{fig:sub1}, players $2$ and $3$ can apply the observables $B_{x \wedge y}$ and $C_z$ on a shared Bell state $\ket{\psi}_{23}$ while player $1$ always outputs $0$. The optimal observables are given by
\begin{align*}
    B_0&=\frac{1}{\sqrt{5}}X+\frac{2}{\sqrt{5}}Z, \quad &&C_0=\frac{1}{\sqrt{2}}(X+Z),\\
    B_1&=\frac{2}{\sqrt{5}}X-\frac{1}{\sqrt{5}}Z, \quad &&C_1=\frac{1}{5\sqrt{2}}X+\frac{7}{5\sqrt{2}}Z,
\end{align*}
where $X$ and $Z$ are Pauli operators. This implies that we have
\begin{align*}
p(1bc|xyz)&=p(1|x)p(bc|xyz)=0  \textrm{ for all } b,c,x,y,z \in \{0,1\} \textrm{ and}   \\
    p(000|xyz)&=\langle \psi | B_{x \wedge y}^+ \otimes C_z^+ | \psi \rangle\\
    p(001|xyz)&=\langle \psi | B_{x \wedge y}^+ \otimes C_z^- | \psi \rangle\\
    p(010|xyz)&=\langle \psi | B_{x \wedge y}^- \otimes C_z^+ | \psi \rangle\\
    p(011|xyz)&=\langle \psi | B_{x \wedge y}^- \otimes C_z^- | \psi \rangle
\end{align*}
where $B_{x \wedge y}^+, C_z^+$ and $B_{x \wedge y}^-, C_z^-$ are projectors corresponding to outputs $0$ and $1$ respectively. 
This results in a winning probability $p_{win}=0.8953$ for the tripartite $AND$ game, which is strictly higher than the value $p_{win}^s=0.875$ achieved in Svetlichny's model~\cite{DetectGMN}. 

Let $\langle a_x b_y c_z \rangle$ be the expected value obtained for inputs $x$, $y$ and $z$. According to Svetlichny \cite{DetectGMN}, a violation of the inequality
\begin{align}
    S_3=\hphantom{+}&\langle a_0 b_0 c_1\rangle+ \langle a_0 b_1 c_0\rangle+ \langle a_1 b_0 c_0\rangle- \langle a_1 b_1 c_1\rangle \nonumber \\
 +& \langle a_1 b_1 c_0\rangle+ \langle a_1 b_0 c_1\rangle+ \langle a_0 b_1 c_1\rangle- \langle a_0 b_0 c_0\rangle \leq 4, \label{eq:svet}
\end{align}
implies the presence of genuine tripartite nonlocality. We can verify that the quantum strategy with $p_{win}=0.8953$ attains the value $S_3=4.4276 >4$, even though we have argued that the strategy does not exhibit genuine tripartite nonlocality.

Bancal \emph{et al.}~\cite{Bancal13} propose a definition of multipartite nonlocality based on the intuition that a given nonlocal correlation between $n$ parties is genuinely $k$-partite nonlocal if at least $k$ parties need to share a genuinely $k$-partite nonlocal resource to simulate it. Things get a bit more nuanced once we consider interconversion between the nonlocal resource shared for simulation~\cite{Barrett05, Ebbe14a}, i.e., quantum entangled state, nonlocal box, communication channel etc. Bancal \emph{et al.}~\cite{Bancal09, DetectGMN} have further developed the idea by quantifying the amount of multipartite nonlocality and generalizing it to arbitrary dimensions. Curchod \emph{et al.}~\cite{Curchod15} have quantified the notion of the \emph{minimal group size (MGS)} which refers to the smallest number of players that need to share a given type of nonlocal resource for simulating a given correlation.

Mao et al.~\cite{Test_22} develop a Bell-type inequality to detect genuine multipartite nonlocality in a network of $n$ players with access to at most $(n-1)$-partite nonlocal resource within the Svetlichny framework. Singh, Sasmal and Pan~\cite{SI_25} use a Sum-of-Squares approach to certify that the quantum strategy that achieves the optimal value for $n$ player Svetlichny inequalities necessarily requires an $n$ partite maximally entangled state. Li and Shang~\cite{measure_22} provide a measure to detect genuinely multipartite entangled states, which can be used to quantify potential quantum resources. Furthermore, the concept of multipartite quantum nonlocality without entanglement has attracted considerable attention. In particular, nonlocality arising from sets of orthogonal product states has been extensively investigated~\cite{OPS_20_1, OPS_25_1, OPS_25_2}.

In this paper, we investigate strategies that take the form of Figure~\ref{fig:trisub} under the notion of Local Operations and Classical Communication with Grouping (LOCCG). Note that even though the model is theory independent, we do not consider strategies utilizing nonlocal boxes in our current work. We begin the next section by providing the definitions needed to construct the LOCCG model and state our main results as two theorems in subsection~\ref{subsec:results}. Section~\ref{sec:reduce} contains our key reduction between multipartite LOCCG and bipartite biased nonlocal games. The proofs for Theorem~\ref{mthm1} and Theorem~\ref{mthm2} are provided in Sections~\ref{sec:mthm1} and~\ref{sec:mthm2} respectively. A fully worked out example to illustrate the reduction process is given in Appendix~\ref{eg:maj}.

\section{Preliminaries}\label{sec:prelim}
In an $n$~party Bell scenario for players $P_1, P_2, \ldots, P_n$, the players are spatially separated and no communication is allowed between them once the game starts. Each player $P_i$ receives input bit $x_i$ picked with uniform randomness and outputs bit $a_i$. The players win an $XOR$ game if the parity of their output bits equals a function over the input bits,~i.e.,
\begin{equation*}\label{eq:game}
\bigoplus_{i=1}^n a_i = f(x_1,x_2,...,x_n),
\end{equation*}
where $f:\{0,1\}^n \mapsto \{0,1\}$ is the boolean function the players compute. The function $f$ is known to all players before the game starts. A \textit{multipartite $THRESHOLD$ game} is an XOR game with binary inputs and the parity of the outputs bits should be equal to the boolean threshold function $f_t$ defined as 
\begin{equation}\label{eq:thresh1}
    f_{t} (x_1,x_2,...,x_n)=\begin{cases}
        1 & \sum_{i=1}^n x_i \geq t\\
        0 & \text{otherwise}
    \end{cases} ,
\end{equation}
where $t= \frac{rn}{s}$ represents the threshold such that $r,s \in \mathbb{Z}^+$ and $r \leq s$. Note that the function's output depends only on the Hamming weight $\abs{x}$ of the $n$ input bit string $x \in \{0,1\}^{n}$. The two extremes of the threshold, $t=\frac{n}{2}$ and $t=n$ in Equation (\ref{eq:thresh1}) correspond to the \textit{multipartite $MAJORITY$ game} and \textit{multipartite $AND$ game} respectively. The threshold $t=n$, corresponds to the $AND$ function over the input bits resulting in the winning condition
\begin{equation*}
\bigoplus_{i=1}^n a_i = \bigwedge_{i=1}^n x_i.
\end{equation*} 
For $n=2$, the rule corresponds to the CHSH game~\cite{CHSH69}. The base case for the other extreme $\left(t=\frac{n}{2}\right)$  is the multipartite Majority game for $n=3$. The function $f_{t}$ is then given by
\begin{equation*}
f_{\frac{3}{2}}(x_1,x_2,x_3)=\begin{cases}
    1 & \sum_{i=1}^3 x_i \geq \frac{3}{2}\\
    0 & \text{otherwise,}
\end{cases}
\end{equation*}
resulting in the winning condition $a_1 \oplus a_2 \oplus a_3=f_{\frac{3}{2}}(x_1,x_2,x_3).$ It is sufficient to consider the range $\frac{n}{2} \leq t \leq n$ for the class threshold functions due to the symmetric nature of the game and any value of $t$ between these two extremes corresponds to the general threshold function (\ref{eq:thresh1}).\\
The winning probability and the expected value $V$ of the game are related by $V=2p_{win}-1$, where $n$ is the number of players. The value $V_C$ attained by classical players for a deterministic strategy in a game $f_t$ is given by 
\begin{equation}\label{eq:LOSR}    
V_C=\frac{1}{2^n}\sum_{x \in \{0,1\}^n } (-1)^{f_t(x)} \prod_{i=1}^n \hat{a}_{x_i}^i ,
\end{equation}
where $x=x_1 x_2 \ldots x_n$ is the length $n$ input bit string and $\hat{a}_{x_i}^i=(-1)^{a_{x_i}^i}$ is the output of the $i^\textrm{th}$ player mapped to $\{\pm 1\}$ on input $x_i$. Similarly, the quantum value $V_Q$ for the same game is given by 
\begin{equation}\label{eq:LOE}
V_Q= \frac{1}{2^n}\sum_{x \in \{0,1\}^n} (-1)^{f_t(x)}\bra{\psi}A_{x_1}^1 \otimes A_{x_2}^2 \otimes \ldots \otimes A_{x_n}^n\ket{\psi},
\end{equation}
where $|\psi \rangle$ is the shared entangled state and $A^i_{x_i}$ is the binary observable of $i^{\textrm{th}}$ player on input $x_i$. 

 \begin{figure}[t]
    \centering
    \includegraphics[width=0.6\linewidth]{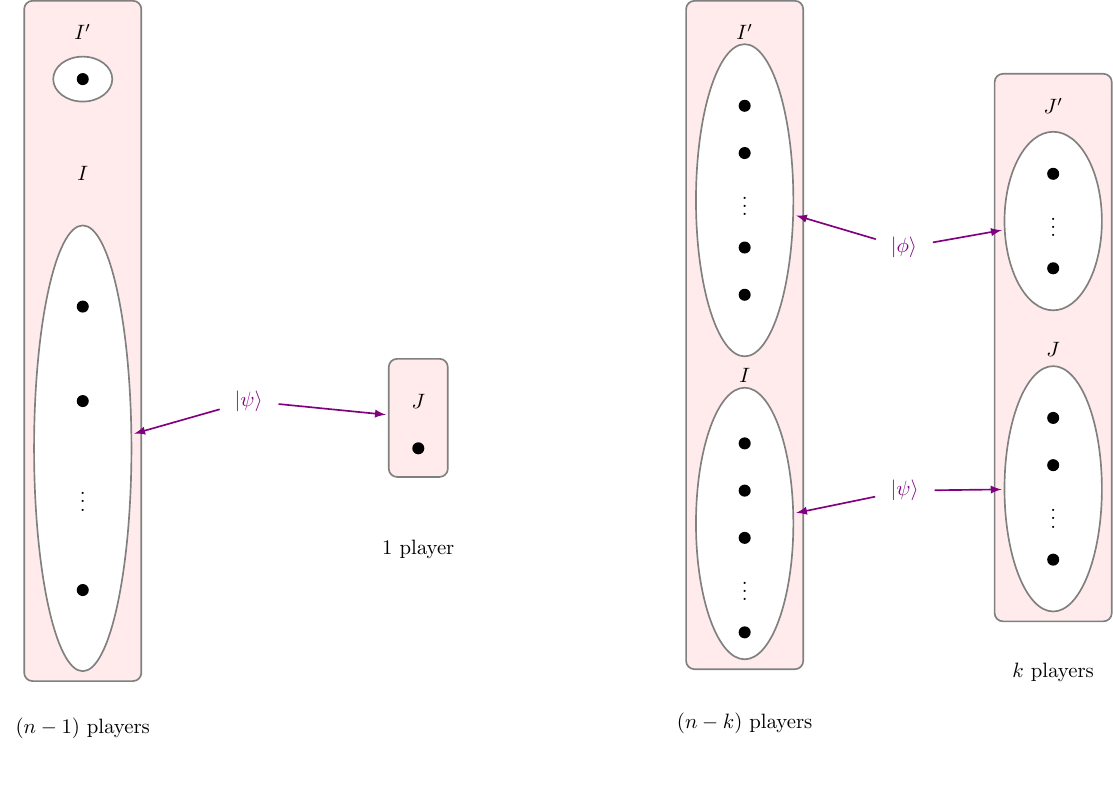}
    \caption{Different groupings of the LOCCG model using communication and shared entanglement as resource. On the left we have grouping $(n-1,1)$ with communication allowed between $n-1$ players in the set $I \cup I'$ and shared entanglement across the $n-1$ players in the set $I \cup J$. Note that both the sets $I'$ and $J$ in this case contain a single player. Figure on the right shows general grouping $(n-k,k)$, with communication allowed only between $n-k$ players in the set $I \cup I'$ and $k$ players in the set $J \cup J'$. At the same time, $n-k$ players in the set $I \cup J$ share an entangled state $\ket{\psi}$, while $k$ players in the set $I' \cup J'$ share an entangled state $\ket{\phi}$. The idea being that even though groups can overlap, no resource is shared across a group of size larger than $n-k$. For simplicity we have omitted entanglement in the visualization that may be shared within each group $A$ and $B$.}
    \label{LoccgDef}
\end{figure}

 Building on our example from Figure~\ref{fig:loccgtri} we now define the LOCCG model. Let the $n$ players be divided into two groups $A$ and $B$ of size $n-k$ and $k$ respectively, given by $(n-k,k)$, where $k \in \{1, 2, \ldots \frac{n}{2}\}$. The binary inputs to players in $A$ and $B$ are given by $x=x_1 x_2 \ldots x_{n-k}$ and $y=y_1 y_2 \ldots y_{k}$ respectively. Similarly, the binary outputs of each set are given by $a=a_1 a_2 \ldots a_{n-k}$ and $b=b_1 b_2 \ldots b_{k}$. Communication is allowed between players in sets $A$ and $B$, but not across the two sets. Let $I$ and $J$ be the set of players in groups $A$ and $B$ respectively that have access to a nonlocal resource such as entanglement or nonlocal boxes. Since we require that the shared resource  should be at most $n-k$ partite, the conditions $|I|\neq 0$, $|J|\neq 0$ and $|I| + |J| \leq  n-k$  are imposed on sets $I$ and $J$. Let $I'$ and $J'$ denote complements of sets $I$ and $J$ as shown in Figure \ref{LoccgDef}. Similarly, the players in these sets may share a nonlocal resource that is at most $n-k$-partite. We denote by $s_I$ a binary string of length $|I|$.

\begin{defn}\label{def:multilocc}
A distribution $p(ab|xy)$ generated by $n$ players in an LOCCG setting is considered $(n-k)$-partite if it admits the decomposition 
\begin{equation}
   p(ab|xy)=\sum_{\lambda, a_I, b_J} q_{\lambda} \, p_{\lambda} (a_I b_J \, | \, xy ) \, p_{\lambda} ( a_{I'} b_{J'}  \,| \, xy ).
\end{equation}
\end{defn}
A distribution that does not satisfy Definition~\ref{def:multilocc} requires a resource larger than $(n-k)$-partite. For example, fixing $k=1$ results in distributions attainable via at max $(n-1) $-partite nonlocal resources. Assuming we have fixed the set $A$ of size $n-1$, the explicit decomposition for the resulting LOCCG distribution is given by
\begin{align}
    p(a_1 a_2 \ldots a_{n-1} b \, | \, xy) = \hphantom{+}& \sum_{\lambda_1} q_{\lambda_1} \, 
    p_{\lambda_1}(a_1\, | \, x) \, p_{\lambda_1}(a_2 a_3 \ldots a_{n-1}b \, | \, x y) \nonumber \\
    +& \sum_{\lambda_2} q_{\lambda_2} \, p_{\lambda_2}(a_2 \, | \,x) \, p_{\lambda_2}(a_1 a_3 \ldots a_{n-1}b \, | \,x y) \nonumber \\
    &  \makebox[\widthof{$\sum_{\lambda_2} q_{\lambda_2} \, p_{\lambda_2}(a_2\, | \,x_2) \, p_{\lambda_2}(a_1 a_3 \ldots a_{n-1}b \, | \, x_1 x_2 \ldots x_n y)$}][c]{\vdots} \nonumber
    \\   
    +& \sum_{\lambda_{n-1}} q_{\lambda_{n-1}} \, p_{\lambda_{n-1}}(a_{n-1} \, | \, x) \, p_{\lambda_{n-1}}(a_1 a_2 \dots a_{n-2}b \, | \, x y),
\end{align}
where the input string $x = x_1 x_2 \ldots x_{n-1}$. A similar argument applies to $(n-k)$-partite correlations for $k>1$. We associate the optimal classical and quantum values with $V_C^*$ and $V_Q^*$, respectively. In the case of $V_C^*$, the players are grouped into communicating sets $A$ and $B$, with no additional nonlocal resources.

\subsection{Our Results}\label{subsec:results}

\begin{figure}
\begin{center}
\includegraphics[width = \linewidth/2]{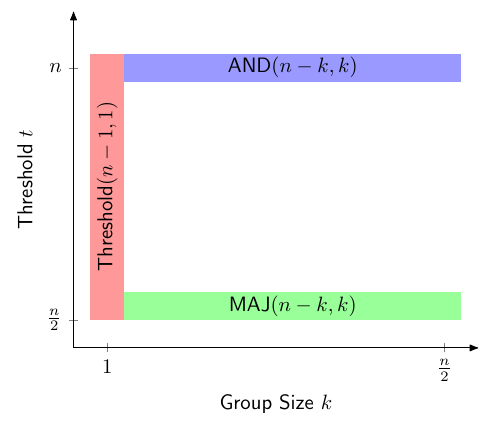}
\end{center}
\caption{\label{fig:main} Visual summary of the games in our results. Quantum advantage exists for $THRESHOLD(n-1,1)$, while there is no advantage for $AND(n-k,k)$. We show that quantum advantage exists for choices of $MAJORITY(n-k,k)$, but it is open to determine how the advantage behaves as we scale $k$ in terms of $n$.}
\end{figure}

We state our main results regarding multipartite nonlocality bounds for $THRESHOLD$ games in Theorems~\ref{mthm1} and~\ref{mthm2}. For our current work, we restrict the nonlocal resource shared across the groups to be a quantum resource. The class of functions to which our results apply are visually summarized in Figure~\ref{fig:main}. 
\begin{theorem}\label{mthm1}
The LOCCG values for  $n$ player $THRESHOLD$ games with $\frac{n}{2} \leq  t \leq n$ are given by the following bounds.

\begin{enumerate}[label=(\alph*), ref=\ref{mthm1}\alph*]
    \item $THRESHOLD$ games with grouping $(n-1,1)$ have the optimal classical value $V^*_C$ given by 
    \begin{align*}
        V^*_C= 1-\frac{\alpha}{2^{n-1}}.
\end{align*}
The quantum value for the same class of games is bounded by the optimal value $V_Q^*$ given by
\begin{equation*}
        V_Q^*= \sqrt{2}\sqrt{(2^{n-2}-\alpha)^2+(2^{n-2})^2} / 2^{n-1},
\end{equation*}
where
\begin{align*}
       \alpha & =\begin{cases}
    \binom{n-1}{\lfloor t\rfloor} &  \textrm{if } n\Mod{s} \neq 0 \\
    \frac{r}{s-r} \binom{n-1}{\lfloor t \rfloor} &  \textrm{if } n \Mod{s} = 0 \\
    1 & \textrm{if } r = s.
\end{cases}
\end{align*} \label{mthmA}
    \item There is no quantum advantage for the $n$ player $AND$ game $(t=n)$ with general $(n-k,k)$ grouping, where $k \neq 1.$  The bound $V^*_Q$ equals $V^*_C$ and is given by
    \begin{equation*}
        V^*_C=V^*_Q = 1-\frac{1}{2^{n-1}}.
    \end{equation*}
    The quantum advantage for $k=1$ is established in Theorem (\ref{mthmA}).\label{mthmB}
\end{enumerate}
\end{theorem}
Note that the quantum advantage in Theorem (\ref{mthmA}) disappears as $n$ becomes large.
\begin{corollary}
The quantum advantage in  Theorem (\ref{mthmA}) disappears in the limit of $n$~i.e.,
\begin{equation*} 
\lim_{n \to \infty} {\frac{V^*_{Q}}{V^*_{C}}} =1.
\end{equation*}
\end{corollary}
We prove Theorem~\ref{mthm1} in Section~\ref{sec:mthm1} by showing a reduction to biased nonlocal games introduced by Lawson, Linden and Popescu~\cite{lawson10}. The bias here refers to the criteria that we no longer assume a uniform distribution for the players' inputs. For example, in the case of CHSH we now have probabilities $p_x$ and $p_y$ for binary inputs $x$ and $y$ to players $1$ and $2$ respectively. Analogous to the definition of the biased CHSH game in~\cite{lawson10}, we present a general definition of \textit{bipartite biased nonlocal games} for integer inputs and binary outputs.

\begin{defn}\label{biased_def}
A \textit{bipartite biased nonlocal game} is a two player XOR game with integer inputs $\hat{x} \in \{0,1,\ldots, d_1\}$ and $\hat{y} \in \{0,1,\ldots, d_2\}$ chosen via an arbitrary probability distribution $p(\hat{x},\hat{y})$. The winning criteria of the game is defined as 
\begin{equation*}
a \oplus b = f(\hat{x},\hat{y})
\end{equation*}
where the outputs $a,b$ of the two players are binary and the function $f(\hat{x},\hat{y})$ is a boolean function $f: \{\hat{x},\hat{y}\} \to \{0,1\}$ defined on the integer inputs.
\end{defn}  

We show that calculating the bound on multipartite LOCCG games reduces to a version of the biased games in Section \ref{sec:reduce}. The reduced game for $MAJORITY$ in Theorem~\ref{mthm2} corresponds to a bipartite game with integer inputs. 
\begin{theorem}\label{mthm2}
The LOCCG classical value of the  $n$ player $MAJORITY$ game $\left( t=\frac{n}{2} \right)$ with $(n-k,k)$ grouping, $k \in \{2, \dots, \frac{n}{2} \}$ is given by the bound
\begin{equation*}
    V^*_C= 1- \frac{\gamma}{2^{n-1}},
\end{equation*}
where the factor $\gamma$ is given by
\begin{equation*}
\gamma=\begin{cases}
    \displaystyle\sum_{j=0}^{\frac{k}{2}-1} \displaystyle\sum_{i=\tilde{t}-k+j}^{\tilde{t}-1-j} w_j \mu_{i} & \textrm{if } k \textrm{ is even, } \\
   \displaystyle\sum_{j=0}^{\frac{k-1}{2}} \displaystyle\sum_{i=\tilde{t}-k+j}^{\tilde{t}-1-j} w_j \mu_{i} & \textrm{if } k \textrm{ is odd.}
\end{cases}
\end{equation*}
where
$ \tilde{t}=\lceil \frac{n}{2} \rceil, 
\mu_i=\binom{n-k}{i},$ and $ \omega_j=\binom{k}{j}.$ 
The optimal quantum protocols for $n \in \{4, \ldots ,10 \}$ with $k=2$,~i.e.,$(n-2,2)$ are given in Table \ref{tab3}.
\end{theorem}
Even though the optimal solution for all $THRESHOLD$ games can be obtained numerically via our SDP (see Appendix \ref{SDP}), we do not have a closed form solution for $MAJORITY$. We observe that while the LOCCG optimal quantum value is greater than the optimal classical value i.e. $V^*_Q > V^*_C$ for fixed values of $n$ and $k$, the quantum advantage disappears in the limit of $n$ for a fixed $k$.
\begin{equation*}
    \lim_{n \to \infty} \frac{{V^*_Q}}{{V^*_{C}}} =1 .
\end{equation*}
On the other hand, for $k=\frac{n}{2}$, the quantum advantage seems to persist even in the limit,~\mbox{i.e.,}\,
    \begin{equation*}
\lim_{n \to \infty}\frac{{V^*_{Q}}}{{V^*_{C}}} > 1 .
    \end{equation*}

\section{Reducing LOCCG to Biased Games}
\label{sec:reduce}
\begin{figure*}[t]
\centering
   \includegraphics[width=\linewidth]{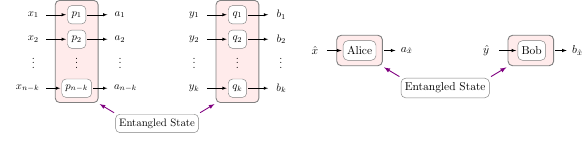}
\caption{The LOCCG model is shown on the left and the reduced biased game is on the right.}
\label{fig:models}
\end{figure*}
We reduce the LOCCG model to bipartite biased games. Since communication is allowed within each group, WLOG we consider that all parties broadcast their input to one chosen player within the group. This allows us to pick one player from each group, let us say Alice and Bob, who know the entire input strings $x \in \{0,1\}^{n-k}$ and $y \in \{0,1\}^k$ of their groups, respectively. 

Let every player in group $A$ share EPR pairs with Alice and similarly for players in $B$ with Bob. Using these EPR pairs, the players in $A$ and $B$ can simply teleport their share of the state to Alice and Bob respectively. Now, Alice and Bob can perform a joint operation on the entirety of their group's share of the entangled state. So, rather than having a tensor product structure over the observables $A_{x_i}^i$ of the $n$ players, Alice and Bob apply observables $D_x$ and $C_y$ on their group's share of the entangled state and output bits $a_x$ and $b_y$ respectively (Figure~\ref{fig:models}).

This allows for the possibility that Alice and Bob use an entangled state that is larger than $(n-k)$-partite. However, for $THRESHOLD$ games we show that Alice and Bob can attain the optimal value using measurements on an EPR state. The classical and quantum values are now given by
 \begin{align}
     V_C &=\frac{1}{2^n} \sum_{x \in \{0,1\}^{n-k}} \sum_{y \in \{0,1\}^k} (-1)^{f_{t}(x,y)}  \hat{a}_{x} \hat{b}_{y} \textrm{ and} \label{CVal} \\
     V_Q &=\frac{1}{2^n} \sum_{x,y} (-1)^{f_{t}(x,y)}\bra{\phi} D_x \otimes C_y \ket{\phi} \label{QVal},
 \end{align}
where $\ket{\phi}$ is any entangled state shared between Alice and Bob. Recall that the threshold function $f_t(x,y)$ is given by
\begin{equation}\label{criteria}
f_t(x,y)=\begin{cases}
    1 & \textrm{if } |x| + |y| \geq t \textrm{ and} \\
    0 & \textrm{otherwise.}
\end{cases}
\end{equation}

We group the input strings for Alice and Bob in terms of their Hamming weight, which gives us $\abs{x} \in \{0,1,\ldots,n-k\}$ and $\abs{y} \in \{0,1, \ldots,k\}$. Observe that based on her input, there exist two cases where Alice has enough information to determine the winning output. 
\begin{enumerate}
    \item If Alice's weight satisfies $0 \leq |x| \leq \tilde{t}-k-1,$ where $\tilde{t}=\lceil \frac{rn}{s} \rceil$ then even if Bob's input weight $\abs{y}=k$,~i.e., all ones input, the function $f_t(x,y)$ evaluates to $0$.  
    \item Assuming $t \leq n-k$, then if Alice's input weight lies in the range $\tilde{t} \leq |x| \leq n-k$, she already knows that the function $f_t(x,y)$ evaluates to $1$.
\end{enumerate}
We define the variable $\hat{x}$ as Alice's input which is obtained by shifting $\abs{x}$, such that
\begin{equation}\label{hatinput}
    \hat{x}=\begin{cases}
    |x|-\tilde{t}+k+1 & \textrm{if } \tilde{t}-k \leq |x| \leq \beta \\
    0 & \textrm{otherwise,}
\end{cases}
\end{equation}
where $\beta$ is given by 
\begin{equation}\label{beta}
    \beta=\begin{cases}
    \tilde{t}-1 & \textrm{if } t \leq n-k \textrm{ and } \\
    n-k & \textrm{if } t > n-k.
\end{cases}
\end{equation}
In the first case for $\beta$, Alice's input $\hat{x}$ now ranges between $0 \leq \hat{x} \leq k$, while in the second case we have $0 \leq \hat{x} \leq n-\tilde{t}+1$. Bob's input case is simpler; we have $\hat{y}=\abs{y}$ with $0 \leq \hat{y} \leq k.$ Finally, we restate the winning criteria in terms of a modified function $g(\hat{x},\hat{y})$, given by
\begin{equation}\label{Bcriteria}
g(\hat{x},\hat{y})=\begin{cases}
    1 &  \textrm{if } \hat{x} + \hat{y} >  k, \textrm{ and}\\
    0 & \textrm{otherwise.}
\end{cases}
\end{equation}

\begin{figure}[t]
    \centering
    \includegraphics{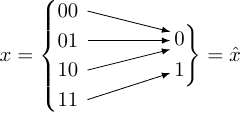}
    \caption{Input mapping of LOCCG $(2,1)$ group $A$'s input $x$ to a biased CHSH game with $n=t=3$.}
    \label{red_eg}
\end{figure}

For example, Figure~\ref{red_eg} shows the input mapping for the example discussed in Section~\ref{int}. The LOCCG $(2,1)$ game with $n=t=3$ is reduced to a biased CHSH game. The input $x \in \{00, 01, 10\}$ of group $A$ is mapped to Alice's input bit $\hat{x}=0$, since the $AND$ of the inputs must evaluate to $0$ regardless of the value of $y$. In the other case, when input $x=11$, the $AND$ of the inputs equals $y$. 

The new input distributions $p_{\hat{x}}$ and $p_{\hat{y}}$ over $\hat{x}$ and $\hat{y}$  are given by
\begin{align*}
p_{\hat{x}} & = \frac{\nu_{\hat{x}}}{2^{n-k}}, 
&q_{\hat{y}} & = \frac{w_{\hat{y}}}{2^k}, \textrm{ where} \\
\nu_0&= 2^{n-k}-\sum_{i=\tilde{t}-k}^\beta \mu_i \textrm{ and}&& \\
    \nu_{\hat{x}} &=\mu_{\tilde{t}-k-1+\hat{x}} \textrm{ for } 1 \leq \hat{x} \leq n-\beta'. &&
\end{align*}
where, 
$$\beta'=\begin{cases}
    n-k & t \leq n-k\\
    \tilde{t}-1 & t > n-k 
\end{cases}.$$
The expressions for the classical and quantum value of the resulting biased game are given by
    \begin{align}
        V_{C}&=\sum_{\hat{x}} \sum_{\hat{y}} (-1)^{g(\hat{x},\hat{y})} p_{\hat{x}} q_{\hat{y}} \hat{a}_{\hat{x}} \hat{b}_{\hat{y}} \textrm{ and} \label{BiasedCVal}\\
        V_{Q}&=\sum_{\hat{x},\hat{y}} (-1)^{g(\hat{x},\hat{y})} p_{\hat{x}} q_{\hat{y}} \bra{\phi } \hat{D}_{\hat{x}} \otimes \hat{C}_{\hat{y}} \ket{\phi}. \label{BiasedQVal}
    \end{align}
This completes our construction of the bipartite biased game. To better illustrate the details, we construct an example reduction along with the player strategy for $MAJORITY$ in Appendix~\ref{eg:maj}. 

The reduction can indeed be applied to multipartite non-XOR games as well. However, the construction along with the winning criterion, input dimension and distribution of the resulting biased bipartite game need to be tailored to reflect the new structure. 

\section{Proof of Theorem~\ref{mthm1}}
\label{sec:mthm1}
\begin{figure}
    \centering
    \includegraphics[width=\linewidth]{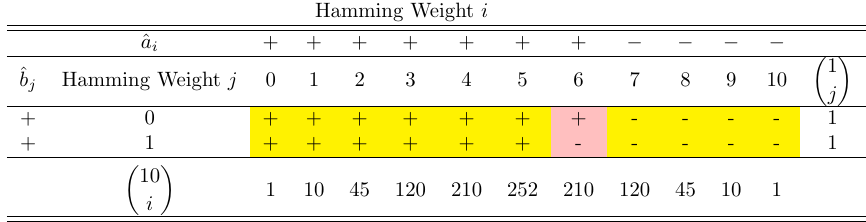}
    \caption{A visual description of the optimal classical strategy for $n=11$, $t=\frac{7n}{12} \approx 6.4$ and $\tilde{t}=7$ with grouping $(n-1,1) = (10,1)$. The row and column; $\hat{a}_i$ and $\hat{b}_j$ represent the binary output of Alice and Bob. The indices $i$ and $j$ represent the Hamming weight of the input strings. The $+$ and $-$ entries in each cell represent evaluation of $(-1)^{f_t}$ for the threshold function $f_t$ (\ref{eq:thresh1}). The yellow columns correspond to the interval $0 \leq |x| \leq \tilde{t} -2$ and $\tilde{t} \leq |x| \leq n-1.$  The entries in yellow cells contribute positively to the expected value while the entries in the pink cells cancel each other. The optimal classical value for $(10,1)$ in LOCCG model is $V_C^*=1-\frac{210}{2^{10}} \approx 0.794.$ The yellow columns represent the case where the function is evaluated with certainty by the group and they map onto the input $\hat{x}=0$ whereas the pink column maps onto the input $\hat{x}=1$  for Alice in the biased CHSH game.}
    \label{tab4}
\end{figure}
We prove Theorem (\ref{mthmA}) by considering the corresponding reduced biased game, for which $\hat{x}, \hat{y} \in \{ 0,1 \}$ and the winning criteria $g(\hat{x},\hat{y})$ is given by 
\begin{equation*}
g(\hat{x},\hat{y})=\begin{cases}
    1 &  \textrm{if } \hat{x} + \hat{y} >  1, \textrm{ and}\\
    0 & \textrm{otherwise.}
\end{cases}
\end{equation*}

The new game is a biased CHSH game with Alice's bias $p_{\hat{x}}$ and Bob's bias $p_{\hat{y}}$ given by
\begin{align*}
p_0 & = 1 - p_1,  \, p_1 = \frac{\alpha}{2^{n-1}}  \textrm{ and}\\
q_0 & = \frac{1}{2} = q_1.
\end{align*}
Recall that the classical value $V_C$ for biased CHSH may be expressed as 
\begin{equation*}
V_{C}= p_0 q_0 \hat{a}_0 \hat{b}_0 + p_0 q_1 \hat{a}_0 \hat{b}_1 + p_1 q_0 \hat{a}_1 \hat{b}_0 - p_1 q_1 \hat{a}_1 \hat{b}_1.    
\end{equation*}
If Alice and Bob always output $\hat{a}_{\hat{x}} = \hat{b}_{\hat{y}} = +1$, they achieve the optimal value
\begin{equation*}
V^*_{C} = 1-\frac{\alpha}{2^{n-1}}.
\end{equation*}
The table in Figure~\ref{tab4} represents the optimal classical strategy of the $11$ partite $THRESHOLD$ game with grouping $(10,1)$ for threshold $t=\frac{7n}{12}$ and its reduction to the biased CHSH game.
For the quantum strategy, we show that Protocol~\ref{prot1} achieves $V^*_{Q}$.
\begin{protocol}\label{prot1}
Let $|\psi\rangle$ be the EPR state shared by Alice and Bob,~i.e. $
\ket{\psi} = \frac{1}{\sqrt{2}}\left(\ket{00} + \ket{11} \right)$ and let $X$ and $Z$ be the Pauli matrices. The observables $\hat{D}_{\hat{x}}$ and $\hat{C}_{\hat{y}}$ for Alice and Bob respectively are given by
\begin{align*}
   \hat{D}_0 & = \frac{(2^{n-2}-\alpha) X + 2^{n-2} Z}{\sqrt{\alpha^2-2^{n-1}\alpha+2^{2n-3}}} ,\\
    \hat{D}_1 & =  \frac{2^{n-2} X - (2^{n-2}-\alpha)Z}{\sqrt{\alpha^2-2^{n-1}\alpha+2^{2n-3}}}\\
    \hat{C}_0&=\frac{1}{\sqrt{2}} (X+Z),\\
    \hat{C}_1&=\frac{(\alpha^2-2^n\alpha+2^{2n-3})X + (2^{2n-3}-\alpha^2) Z}{\sqrt{2}(\alpha^2-2^{n-1}\alpha+2^{2n-3})}.
\end{align*}
\end{protocol}

\begin{figure}[t]
\centering
   \includegraphics[width=\linewidth/2]{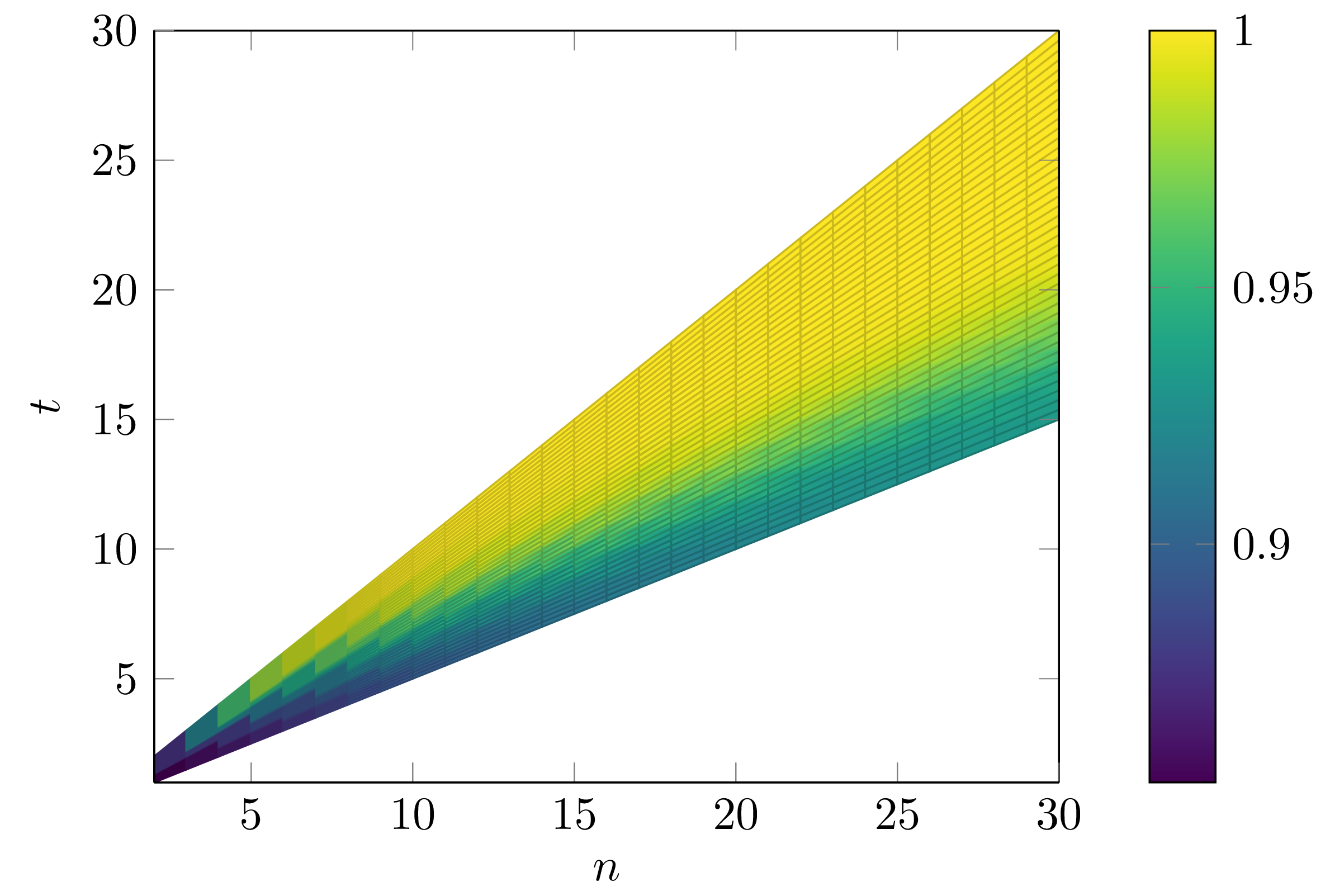}
\caption{The winning probability for Protocol~\ref{prot1} with the number of players $n$ on the horizontal axis and the game threshold $t$ on the vertical axis.}
\label{fig:mapn}
\end{figure}

Figure~\ref{fig:mapn} shows the winning probability attained by Protocol~\ref{prot1} as we vary the number of players $n$ and the game threshold $t$. The optimality of Protocol~\ref{prot1} follows from the biased CHSH game of Lawson, Linden and Popescu~\cite{lawson10}. Here, we provide an alternate proof using the SDP formulation from Appendix \ref{SDP} to make explicit dependency on $n$.
\begin{lemma}\label{lm1}
Protocol~\ref{prot1} is optimal for the biased CHSH game.
\end{lemma}
\begin{proofof}{Lemma~\ref{lm1}}
The value of Protocol \ref{prot1} may be determined by first calculating the expected values
\begin{align*}
    \bra{\psi} \hat{D}_0 \otimes \hat{C}_0 \ket{\psi} &= \bra{\psi} \hat{D}_0 \otimes \hat{C}_1 \ket{\psi}\\
    &=\frac{2^{n-1}-\alpha}{\sqrt{2}\sqrt{(2^{n-2}-\alpha)^2+(2^{n-2})^2}},
    \\
    \bra{\psi} \hat{D}_1 \otimes \hat{C}_0 \ket{\psi} &=- \bra{\psi} \hat{D}_1 \otimes \hat{C}_1 \ket{\psi}\\
    &=\frac{\alpha}{\sqrt{2}\sqrt{(2^{n-2}-\alpha)^2+(2^{n-2})^2}}.
\end{align*}
Plugging these values into our expectation expression 
\begin{equation*}
\langle p_0q_0 \hat{D}_0 \hat{C}_0 + p_0q_1 \hat{D}_0 \hat{C}_1 +p_1q_0 \hat{D}_1  \hat{C}_0-p_1q_1 \hat{D}_1  \hat{C}_1 \rangle,
\end{equation*}
yields the optimal value
\begin{equation*}
V^*_Q = \sqrt{2} \sqrt{(2^{n-2}-\alpha)^2+(2^{n-2})^2}/2^{n-1}.
\end{equation*}
The symmetric weight matrix $W$ in the SDP primal is given by 
\begin{equation*}
    W=\begin{bmatrix}
        0 & 0 & p_0 q_0 & p_0 q_1\\
        0 & 0 & p_1 q_0 & -p_1 q_1\\
        p_0 q_0 &  p_1 q_0 & 0 & 0\\
        p_0 q_1 & -p_1 q_1 & 0 &0  
    \end{bmatrix}
\end{equation*} 
while the dual vector $b$ is the all one vector of size $4$. For any entangled state $\ket{\phi}$ shared between the players Alice and Bob, the vectors $\{u_x,u_y,v_0,v_1\}$ are given by
\begin{align*}
    u_0 &= (\hat{D}_0 \otimes \mathbb{I}) \ket{\phi},  \\
    u_1 &= (\hat{D}_1 \otimes \mathbb{I}) \ket{\phi},  \\
    v_0 &= (\mathbb{I}\otimes \hat{C}_0) \ket{\phi} \textrm{ and}  \\
    v_1 &= (\mathbb{I} \otimes \hat{C}_1) \ket{\phi}.
\end{align*}
Thus, we have $\langle \phi | \hat{D}_i \otimes \hat{C}_j | \phi \rangle= u_i \cdot v_j.$
Protocol~\ref{prot1} serves as a valid feasible solution of the primal. A solution to the Dual is given by the vector $\lambda \in \mathbb{R}^4$, where
\begin{align*}
   \lambda_1&=\frac{(2^{n-1}-\alpha)^2}{2^{n-1}\sqrt{2}\sqrt{(2^{n-2}-\alpha)^2+(2^{n-2})^2)}},
    \\
    \lambda_2&=\frac{\alpha^2}{2^{n-1}\sqrt{2}\sqrt{(2^{n-2}-\alpha)^2+(2^{n-2})^2)}},\\
\lambda_3&=\frac{\alpha^2 +(2^{n-1}-\alpha)^2}{2^n\sqrt{2}\sqrt{(2^{n-2}-\alpha)^2+(2^{n-2})^2)}} =\lambda_4.
\end{align*}
We can verify that the dual solution matches the primal value by calculating the summation
\begin{equation*}
\sum_{i=1}^4 \lambda_i=\sqrt{2}\sqrt{(2^{n-2}-\alpha)^2+(2^{n-2})^2} / 2^{n-1}.
\end{equation*}
The characteristic polynomial of the sdp dual matrix $K=2 \textrm{diag}(\lambda)-W$ is given by
\begin{equation*}
c_4 x^4+c_3 x^3 + c_2 x^2 + c_1 x +c_0,    
\end{equation*}
where $\textrm{diag}(\lambda)$ is a diagonal matrix with the vector $\lambda$ across the diagonal and the coefficients $c_i$ have the form
\begin{align*}
    c_4&=1, \\
    c_3&=-\frac{\sqrt{2^{2n}-\alpha(2^{n+2}-2^3 \alpha)}}{2^{n-2}}, \\
    c_2&=\frac{\alpha(2^{n+2}-2^3 \alpha)}{2^{2n}-\alpha(2^{n+2}-2^3 \alpha)}+\frac{11 \alpha^2}{2^{2n-2}}+\frac{9}{2}-\frac{11\alpha}{2^{n-1}}, \\
    c_1 &=- \frac{(2^{4n}-8\alpha(2^{n}-2 \alpha)(2^{2n}-4\alpha(2^n-2\alpha)))}{2^{15n+5}\sqrt{2^{2n}-\alpha(2^{n+2}-2^3\alpha)}}, \\
    c_0 &=\frac{9\alpha^2(2^n-2\alpha)^2}{2^{4n}}. 
\end{align*}
The coefficients $c_4$ and $c_0$ are always positive for all values of $\alpha$. The coefficient $c_3$ is negative for all values of $\alpha$ as well, while for $c_2$ and $c_1$, we have the following inequalities
\begin{align*}
    \frac{\alpha(2^{n+2}-2^3 \alpha)}{2^{2n}-\alpha(2^{n+2}-2^3 \alpha)}+\frac{9}{2}&>\frac{11\alpha(2^{n-1}-\alpha)}{2^{2n-2}},\\
    2^{4n}+2^5 \alpha^2 (2^{n}-2\alpha)^2 &>2^{2n+3}\alpha(2^{2n}-2\alpha).
\end{align*}
We can verify that the inequalities are valid for all values of $n$ and $\alpha.$ Thus, the coefficients $c_2$ and $c_1$ have alternating signs,~i.e., $c_2>0$ and $c_1<0.$ Based on the alternating signs of coefficients of the characteristic polynomial, $c_i \cdot c_{i+1} <0$ for $0 \leq i < 4$, the matrix $K$ is a positive semidefinite matrix (Corollary 7.2.4 in \cite{MatA}). Hence, we have shown that Protocol~\ref{prot1} is optimal.
\end{proofof}
This concludes the proof of Theorem~(\ref{mthmA}). For the proof of Theorem~(\ref{mthmB}) note that even when we consider the general $(n-k,k)$ grouping, the reduced game is still a biased CHSH game. The binary input probabilities for the biased CHSH game are given by
\begin{equation}\label{prob}
p_0=1-\frac{1}{2^{n-k}} \quad \textrm{and} \quad q_0=1-\frac{1}{2^k},
\end{equation}
where $p_0$ and $q_0$ are the probabilities of Alice and Bob receiving input $0$ respectively. For $2 \leq k \leq \frac{n}{2}$, the probabilities $p_0$ and $q_0$ satisfy the inequality
\begin{equation*}
\frac{1}{2} \leq (2q_0)^{-1} \leq p_0 \leq 1
\end{equation*}
given by Lawson, Linden and Popescu~\cite{lawson10}. These belong to the region $[p_0,q_0]$ where quantum strategies provide no advantage over classical strategies for biased games.

\begin{proofof}{Theorem (\ref{mthmB}).}
We~show that the quantum bound matches the classical bound  i.e. $V_Q^*=V^*_C=1-\frac{1}{2^{n-1}}$ for the biased CHSH game with any grouping other than $(n-1,1)$. The SDP proof for optimality proceeds in a similar fashion except that the new bias $p_{\hat{x}}$ for Alice and $q_{\hat{y}}$ for Bob are given by
\begin{align*}
p_0 & = 1 - p_1,  & p_1 &= \frac{1}{2^{n-k}}  \textrm{ and}\\
q_0 & = 1- q_1, &  q_1 &= \frac{1}{2^k}.
\end{align*}
For the optimal quantum protocol the players measure $\ket{\psi}$ in the computational basis which gives 
\begin{align*}
V^*_Q &= \hphantom{+} p_0 q_0 + p_0 q_1 + p_1 q_0 - p_1 q_1\\
&=\hphantom{+} \left( 1-\frac{1}{2^{n-k}} \right ) \left( 1-\frac{1}{2^k} \right)+\frac{1}{2^k} \left( 1-\frac{1}{2^{n-k}} \right) + \frac{1}{2^{n-k}} \left( 1-\frac{1}{2^k} \right) - \frac{1}{2^n}\\
&= \hphantom{+} 1-\frac{1}{2^{n-1}}.
\end{align*}
The dual solution vector $\lambda$ is given by
\begin{align*}
    \lambda_1&=\frac{1}{2}-\frac{1}{2^{n-k+1}},\\
    \lambda_{2}&=\frac{1}{2^{n-k+1}}-\frac{1}{2^{n}}, \\
    \lambda_{3}&=\frac{1}{2}-\frac{1}{2^{k+1}}, \\
    \lambda_{4}&=\frac{1}{2^{k+1}}-\frac{1}{2^{n}},
\end{align*}
where each $\lambda_{i} \geq 0$ and their summation
\begin{equation*}
\sum_{i=1}^4 \lambda_i = 1-\frac{1}{2^{n-1}},
\end{equation*}
results in a matching solution value. The dual sdp matrix $K$ has the following characteristic polynomial
\begin{equation*}
b_4 x^4+b_3 x^3 + b_2 x^2 + b_1 x +b_0,    
\end{equation*}
where the coefficients $b_i$ are given by
\begin{align*}
     b_4&=1, \\
    b_3&=- (2-\frac{1}{2^{n-2}}),\\
    b_2&=\frac{(2^{n+k}-2^n-2^{2k})(2^n-2^{1+k}+2^{2k})}{2^{2n+2k-5}}, \\
    b_1 &=\frac{4(2^k-1)(2^n-2^k)(2^{2k+1}+2^{n+1}-2^{k+1}-2^{n+k})}{2^{3n+2k}}, \\
    b_0 &=\frac{(2^k-1)(2^{5k+5n}-2^{2k+1}-2^{2n})}{2^{2n+3k}}. \end{align*}
The coefficient $b_4$ is positive and $b_3$ is negative. Since the following inequalities hold for all $1 < k \leq \frac{n}{2}$, the coefficients $b_2$, $b_1$ and $b_0$ have alternating signs i.e. $b_2 >0$, $b_1 < 0$ and $b_0 > 0$. Given the inequalities
\begin{align*}
    2^{n+k} &> 2^n +2^{2k} > 2^{k+1},\\
2^{n+k} + 2^{k+1} &> 2^{n+1} +2^{2k+1} \textrm{ and}\\
2^{5(n+k)}&> 2^{2n} +2^{2k+1},\\
\end{align*}
we have $b_i \cdot b_{i+1} <0$ for $0 \leq i < 4$ and the matrix $K$ is a positive semidefinite matrix (Corollary 7.2.4 in \cite{MatA}). Since the quantum bound matches the classical bound,~i.e.,~$V^*_Q = V^*_C$, the quantum advantage has vanished. 
\end{proofof}

\section{Proof of Theorem~\ref{mthm2}}\label{sec:mthm2}
We begin by constructing the reduced biased game for $MAJORITY$. Since $\tilde{t}=\lceil \frac{n}{2} \rceil$, the inequality $\tilde{t} \leq n-k$ holds for $ 2 \leq k \leq \frac{n}{2}$. This gives us $\beta=\tilde{t}-1$ from Equation~\ref{beta}. The new inputs $\hat{x}$ and $\hat{y}$ for Alice and Bob respectively range between $0 \leq \hat{x}, \hat{y} \leq k$. Alice's input $\hat{x}$ is given by
\begin{align*}
    \hat{x}&=\begin{cases}
        |x|-\tilde{t}+k+1 & \textrm{if } \tilde{t}-k \leq |x| \leq \tilde{t}-1\\
        0 & \textrm{otherwise,}
    \end{cases}
\end{align*}
and Bob's input is $\hat{y}=|y|$. The input probability distributions are given by
\begin{align*}
p_{\hat{x}} & = \frac{\nu_{\hat{x}}}{2^{n-k}}, 
&q_{\hat{y}} & = \frac{w_{\hat{y}}}{2^k}, \textrm{ where} \\
\mu_{\hat{x}} & = \binom{n-k}{\hat{x}}, &w_{\hat{y}}&=\binom{k}{\hat{y}}, \\
\nu_0&= 2^{n-k}-\sum_{i=\tilde{t}-k}^{\tilde{t}-1} \mu_i \textrm{ and}&& \\
    \nu_{\hat{x}} &=\mu_{\hat{x}+\tilde{t}-k-1} \textrm{ for } 1 \leq \hat{x} \leq k. &&
\end{align*}
The winning criteria is determined by Equation~\ref{Bcriteria}. Note that we no longer obtain a biased CHSH game; now the input size is bounded by the group size $k$. Protocol~\ref{prot2}~provides an optimal classical deterministic strategy for the game in its original LOCCG formulation. 

\begin{protocol}\label{prot2}
All members of Bob's group of size $k$, including Bob output $=+1$ for all $y \in \{0,1\}^k.$ Everyone in Alice's group, except Alice outputs $+1$. Alice's output $\hat{a}_x$ is determined by the group size $k$.
\begin{enumerate}
    \item If $k$ is even then
\begin{align*}
    \hat{a}_x=\begin{cases}
        +1 & \textrm{if } 0 \leq |x| \leq \tilde{t}-1-\frac{k}{2}\\
        -1 & \textrm{otherwise}.\\
        \end{cases}
        \end{align*}
    \item If $k$ is odd then
\begin{align*}
    \hat{a}_x=\begin{cases}
        +1 & \textrm{if } 0 \leq |x| \leq \tilde{t}-1-\frac{k-1}{2}\\
        -1 & \textrm{otherwise}.\\
        \end{cases}
\end{align*}
\end{enumerate}
\end{protocol}
Optimality follows from the fact that we maximize the number of inputs for which we can obtain a positive contribution in Equation~\ref{eq:LOSR}. Lemma~\ref{lm2} establishes the value obtained through Protocol~\ref{prot2}.

\begin{figure}
    \centering
    \includegraphics[width=\linewidth]{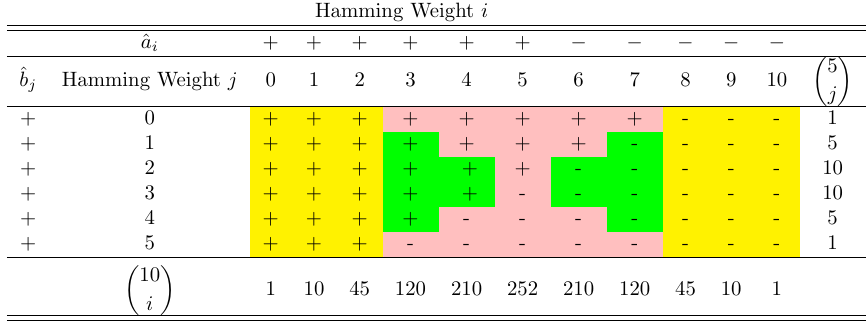}
    \caption{A visual description of the optimal classical strategy for $MAJORITY$ with grouping $(n-k,k) = (10,5)$ and threshold $\tilde{t}=8$. The row and column; $\hat{a}_i$ and $\hat{b}_j$ give the binary output for Alice and Bob. The indices $i$ and $j$ represent the Hamming weight of the input strings. The $+$ and $-$ entries in each cell represent evaluation of $MAJORITY$ on the inputs. The yellow columns correspond to the interval $0 \leq |x| \leq \tilde{t}-k-1$ and $\tilde{t} \leq |x| \leq n-k.$  The entries in yellow cells and green cells are added and contribute positively to the expected value while the entries in pink cells get canceled out. The optimal classical value for $(10,5)$ is $V_C^*=\frac{19184}{2^{15}} \approx 0.5854.$}
    \label{tab1}
\end{figure}

\begin{lemma}\label{lm2}
Protocol~\ref{prot2} achieves the value
\begin{equation*}
    V^*_C= 1- \frac{\gamma}{2^{n-1}},
\end{equation*}
where the factor $\gamma$ is given by
\begin{equation*}
\gamma=\begin{cases}
     \displaystyle\sum_{j=0}^{\frac{k}{2}-1} \displaystyle\sum_{i=\tilde{t}-k+j}^{\tilde{t}-1-j} w_j \mu_{i} & \textrm{if } k \textrm{ is even, } \\
   \displaystyle\sum_{j=0}^{\frac{k-1}{2}} \displaystyle\sum_{i=\tilde{t}-k+j}^{\tilde{t}-1-j} w_j \mu_{i} & \textrm{if } k \textrm{ is odd.}
\end{cases}
\end{equation*}
\end{lemma}

\begin{proofof}{Lemma~\ref{lm2}.}
Recall that the expression for classical value $V_C$ for the $MAJORITY$ function is given by
\begin{equation}\label{eq:maj}
     V_C =\frac{1}{2^n} \sum_{x \in \{0,1\}^{n-k}} \sum_{y \in \{0,1\}^k} (-1)^{MAJ(x,y)}  \hat{a}_{x} \hat{b}_{y}.
 \end{equation}
Note that for all $y \in \{0,1\}^k$, we have
\begin{equation*}
MAJ(x,y)=\begin{cases}
    0 & \textrm{if } 0 \leq |x| \leq \tilde{t}-k-1 \textrm{ and}\\
    1 & \textrm{if } \tilde{t} \leq |x| \leq n-k,
\end{cases}
\end{equation*}
which results in a contribution of 
\begin{equation*}
\frac{1}{2^{n-k}}\left(\sum_{i=0}^{\tilde{t}-k-1} \mu_i +\sum_{i=\tilde{t}}^{n-k} \mu_i \right)
\end{equation*}
to $V^*_C$ after plugging in the values of $\hat{a}_x$ and $\hat{b}_y$. For an even $k$, the strings $\tilde{t}-k \leq |x| \leq \tilde{t}-1$ contribute
\begin{align*}
    &\frac{1}{2^n}\left( \mu_{\tilde{t}-k} + \mu_{\tilde{t}-1}  \right) \left( 2^k -2 w_0 \right) + \frac{1}{2^n}\left( \mu_{\tilde{t}-k+1} + \mu_{\tilde{t}-2}  \right) \left( 2^k -2 (w_0+w_1) \right)\\
    &+ \ldots + \frac{1}{2^n}\left( \mu_{\tilde{t}-\frac{k}{2}-1} + \mu_{\tilde{t}-\frac{k}{2}}  \right) \left( 2^k -2 (w_0+w_1+\ldots + w_{\frac{k}{2}-1}) \right)
\end{align*}
which furthermore generalizes to 
\begin{align*}
    \frac{1}{2^{n-k}} \sum_{i=\tilde{t}-k}^{\tilde{t}-1} \mu_i-\frac{1}{2^{n-1}}\sum_{j=0}^{\frac{k}{2}-1}\sum_{i=\tilde{t}-k+j}^{\tilde{t}-1-j} w_j \mu_i
\end{align*}
By plugging in all the contributions into Equation~\ref{eq:maj}, we obtain
\begin{align*}
V^*_C =&\frac{1}{2^{n-k}}\left(\sum_{i=0}^{\tilde{t}-k-1} \mu_i +\sum_{i=\tilde{t}}^{n-k} \mu_i \right) +   \frac{1}{2^{n-k}} \sum_{i=\tilde{t}-k}^{\tilde{t}-1} \mu_i-\frac{1}{2^{n-1}}\sum_{j=0}^{\frac{k}{2}-1}\sum_{i=\tilde{t}-k+j}^{\tilde{t}-1-j} w_j \mu_i\\
=&1-\frac{1}{2^{n-1}}\sum_{j=0}^{\frac{k}{2}-1}\sum_{i=\tilde{t}-k+j}^{\tilde{t}-1-j} w_j \mu_i
\end{align*} 
Similar evaluation for the strings of length $\tilde{t}-k \leq |x| \leq \tilde{t}-1$ when $k$ is odd gives
\begin{align*}
    &\frac{1}{2^n}\left( \mu_{\tilde{t}-k} + \mu_{\tilde{t}-1}  \right) \left( 2^k -2 w_0 \right) + \frac{1}{2^n}\left( \mu_{\tilde{t}-k+1} + \mu_{\tilde{t}-2}  \right) \left( 2^k -2 (w_0+w_1) \right)+ \ldots  \\
    & +\frac{1}{2^n}\left( \mu_{\tilde{t}-\frac{k+3}{2}} + \mu_{\tilde{t}-\frac{k-1}{2}}  \right) \left( 2^k -2 \sum_{j=0}^{\frac{k-1}{2}-1}w_j \right) + \frac{1}{2^n} \mu_{\tilde{t}-\frac{k+1}{2}} \left( 2^k -2 \sum_{j=0}^{\frac{k-1}{2}}w_j \right)
\end{align*}
which furthermore generalizes to 
\begin{align*}
    \frac{1}{2^{n-k}} \sum_{i=\tilde{t}-k}^{\tilde{t}-1} \mu_i-\frac{1}{2^{n-1}}\sum_{j=0}^{\frac{k-1}{2}}\sum_{i=\tilde{t}-k+j}^{\tilde{t}-1-j} w_j \mu_i
\end{align*}
Plugging in all the contributions for an odd $k$ into Equation \ref{eq:maj} gives 
\begin{align*}
V^*_C =&\frac{1}{2^{n-k}}\left(\sum_{i=0}^{\tilde{t}-k-1} \mu_i +\sum_{i=\tilde{t}}^{n-k} \mu_i \right) +   \frac{1}{2^{n-k}} \sum_{i=\tilde{t}-k}^{\tilde{t}-1} \mu_i-\frac{1}{2^{n-1}}\sum_{j=0}^{\frac{k-1}{2}}\sum_{i=\tilde{t}-k+j}^{\tilde{t}-1-j} w_j \mu_i\\
=&1-\frac{1}{2^{n-1}}\sum_{j=0}^{\frac{k-1}{2}}\sum_{i=\tilde{t}-k+j}^{\tilde{t}-1-j} w_j \mu_i
\end{align*}
This completes the proof of Lemma~\ref{lm2}.
\end{proofof}

As an example, we visualize the optimal classical strategy in the table in Figure~\ref{tab1} for $n=15$ players with grouping $(10,5),$~i.e.,~$k=5$. Table~\ref{tab3} lists the optimal quantum protocols for $n \in \{4, \ldots ,10 \}$ with $k=2$,~i.e.,~$(n-2,2)$. Optimality follows from the provided matching dual solution. This completes our proof of Theorem~\ref{mthm2}.
\begin{center}
    
\begin{table*}[t]
\begin{center}
  
    \begin{tabular}{ccccccccccccc}
    \hline \hline
         $k$  & 2 & 4 & 6 & 8 & 10 & 12 & 14 & 16 & 18 & 20 & 22 & 24 \\
         \hline \\[-8pt]
         $n_k$ & 10 & 26 & 42 & 60 & 78 & 98 & 118 &  140 & 162 & 184 & 208 & 252\\
         \hline \hline
    \end{tabular}
    \end{center}
    \caption{For all $n \geq n_k$, the quantum advantage disappears for $MAJORITY$ with grouping $(n-k,k)$.}
    \label{tab2}
\end{table*}
\end{center}

We note that already for $n=10$ with $k=2$, quantum advantage disappears. For each $n$, we would like to determine a bound in terms $k$ and $t$ where the quantum advantage disappears. Unfortunately, we are unable to provide a closed form solution. Table~\ref{tab2} provides the exact value $n_k$ where $k=2i$ and $1 \leq i \leq 12$, such that the optimal quantum value is strictly higher than the optimal classical value for all $n < n_k$. For odd value of $k$, the optimal quantum value $V^*_Q$ converges to the optimal classical value $V^*_C$ asymptotically in terms of $n$. We formalize this notion via Conjecture~\ref{lm3} which argues that for a grouping $(n-k,k)$ with fixed $k$, the quantum advantage disappears for a large enough $n$.

\begin{conjecture}\label{lm3}
Quantum advantage for $MAJORITY$ with grouping $(n-k,k)$ for a fixed $k$ disappears for $n \geq n_k$.
\end{conjecture}

Since the value $V_C^*$ from Protocol~\ref{prot2} is also attainable by a quantum strategy, we only need to provide a dual solution with matching value to prove Conjecture~\ref{lm3}. The dual solution vector $\lambda$ is given by 
   \begin{equation*}
   \lambda= \left[ 
    \lambda_0,
    \lambda_1,
    \ldots,
    \lambda_k,
    \sigma_0,
    \sigma_1,
    \ldots,
    \sigma_k \right]^T.
   \end{equation*}
For even $k$ the solution entries of $\lambda$ are given by 
\begin{align*} 
    \lambda_0&=\frac{1}{2}-\frac{1}{2^{n-k+1}}\sum_{i=1}^{k} \mu_{\tilde{t}-i},     &&\sigma_{\frac{k}{2}}=\frac{w_{\frac{k}{2}}}{2^{k+1}},\\
    \lambda_i &= \frac{\mu_{\tilde{t}-k+i-1}}{2^{n+1}}\sum_{j=i}^{k-i} w_j, && 1\leq i \leq \frac{k}{2},\\ 
    \lambda_{\frac{k}{2}+i}&=\frac{\mu_{\tilde{t}-\frac{k}{2}+i-1}}{2^{n+1}}\sum_{j=\frac{k}{2}-i+1}^{\frac{k}{2}+i-1} w_j, && 1\leq i \leq \frac{k}{2},\\
    \sigma_i&=\frac{w_i}{2^n}\left( 2^{n-k-1}- \sum_{j=i+1}^{\frac{k}{2}} \mu_{\tilde{t}-j} \right), && 0 \leq i \leq \frac{k}{2}-1,\\
    \sigma_i &=\frac{w_i}{2^n} \left( 2^{n-k-1} - \sum_{j=\frac{k}{2}+1}^{i}\mu_{\tilde{t}-j} \right), && \frac{k}{2}+1  \leq i \leq k.
\end{align*}
We may verify that the sum of these entries gives us the classical bound. Similarly for odd $k$, $\lambda$ solution entries are given by
\begin{align*} 
    \lambda_0&=\frac{1}{2}-\frac{1}{2^{n-k+1}}\sum_{i=1}^{k} \mu_{\tilde{t}-i}, &&    \lambda_{\frac{k+1}{2}}=0, \\
    \lambda_i &=  \frac{\mu_{\tilde{t}-k+i-1}}{2^{n+1}}\sum_{j=i}^{k-i}w_j, && 1\leq i \leq \frac{k-1}{2},\\ 
    \lambda_{i+\frac{k+1}{2}}&= \frac{\mu_{\tilde{t}-\frac{k+1}{2}+i}}{2^{n+1}}\sum_{j=\frac{k-1}{2}-i+1}^{\frac{k+1}{2}+i-1}w_j, && 1\leq i \leq \frac{k-1}{2}.\\
\end{align*}     
Fixing $\delta = 2^{n-k}-\mu_{\tilde{t}-\frac{k+1}{2}}$, the $\sigma$ solution entries may be stated as
\begin{align*}    
        &\sigma_{\frac{k-1}{2}}=\sigma_{\frac{k+1}{2}}=\frac{w_{\frac{k+1}{2}}}{2^{n+1}} \delta,&&\\
    \sigma_i&=\frac{w_i}{2^{n+1}}\left( \delta-2 \sum_{j=i+1}^{\frac{k-1}{2}} \mu_{\tilde{t}-j} \right), && 0 \leq i \leq \frac{k-1}{2}-1,\\
    \sigma_i &=\frac{w_i}{2^{n+1}} \left( \delta-2 \sum_{j=\frac{k+1}{2}+1}^{i}\mu_{\tilde{t}-j} \right), && \frac{k+1}{2}+1  \leq i \leq k.
\end{align*}
We only need to argue that the dual matrix $K$ is positive semidefinite to complete a proof of Conjecture~\ref{lm3}.    \begin{figure}[t]
\begin{center}

\includegraphics[width = \linewidth/2]{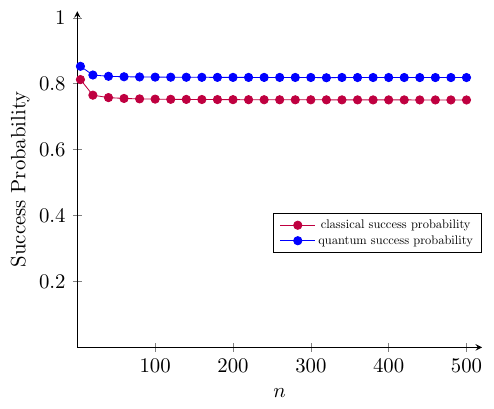}
\end{center}

\caption{\label{fig:maj} Separation between the quantum and classical bounds persists for $MAJORITY$ with grouping $\left( \frac{n}{2},\frac{n}{2} \right)$.}
\end{figure}

On the other hand, numerical evidence suggests that quantum advantage is maintained when the group size $k$ depends on $n$. Figure~\ref{fig:maj} shows an asymptotic separation between quantum and classical bounds for the grouping $(k, k)$, where $k=\frac{n}{2}$. The optimal classical bound is given by
\begin{align*}
V_C^* = \begin{cases}
\frac{1}{2^{n-1}}\left[2^{n-2} + \dbinom{\frac{n}{2}-1}{\frac{n}{4}}^2 \right] & \textrm{if } k \textrm{ is even, and } \\
\frac{1}{2} & \textrm{if } k \textrm{ is odd.}
\end{cases}
\end{align*}
We conjecture that the separation between quantum and classical bounds holds when the group size $k$ depends on $n$ and leave it as an open problem to determine the bound on $k$.

\section{Discussion}\label{sec:discuss}

We have obtained a general procedure to obtain genuine multipartite nonlocality bounds for $THRESHOLD$ games by considering reductions to bipartite biased games. One possible generalization is to identify the larger class of games that correspond to biased games. We may also attempt the reduction in the other direction to better understand measurement dependence. Start with biased games having joint input probability distribution $p(\hat{x}, \hat{y})$ and identify the structure of corresponding multipartite games. Another possibility is to group the $n$ players into $l$ groups and then reduce to $l$-partite biased games.

One motivation for considering genuine multipartite nonlocality bounds is to identify multipartite information principles that limit quantum nonlocality. Gallego \emph{et al.}~\cite{Gallego11} and Yang \emph{et al.}~\cite{Yang12} suggest the need to develop a better understanding of multipartite nonlocal correlations. This naturally necessitates that we first agree on the definition on when a nonlocal correlation is genuinely multipartite. Since games such as \textit{Guess Your Neighbour's Input} are not $XOR$ games, the reduction we use in our construction may need to change.

Our definition of the LOCCG model is operational in the sense that protocols can be realized using physical resources. One generalization of the model may replace shared entanglement across groups with the hypothetical nonlocal boxes. Observe that the resulting biased games can be won with probability one using the appropriate nonlocal box.

\begin{corollary} \label{col1}
Any bipartite biased game obtained from multipartite THRESHOLD game can be won with probability one if Alice and Bob share a nonlocal box defined as
\begin{equation*}
p_{ab|\hat{x}\hat{y}}=\begin{cases}
    \frac{1}{2} & a \oplus b =g(\hat{x},\hat{y}) \\
    0 & \textrm{otherwise,}
\end{cases}    
\end{equation*}
where the function $g(\hat{x},\hat{y})$ is given by equation (\ref{Bcriteria}).
\end{corollary}
Note that Corollary~\ref{col1} may not hold for biased games obtained from multipartite games that are not XOR. 

In the context of an $n$-partite quantum key distribution (QKD) protocol, Xiang has shown that the security of multipartite key distribution is ensured by the monogamy of GHZ correlations~\cite{SI_Crypto_23}. In particular, when an eavesdropper controls the measurement outcomes of up to $n-1$ parties, a violation of Svetlichny’s inequality (SI) guarantees the security of the protocol. However, we have shown that a violation of SI does not necessarily certify genuine multipartite nonlocality. This suggests that we should reassess the security of multipartite QKD within the LOCCG model.

Another direction for future work is to consider LOCCG model involving $k$ groupings of $n$ players with $k > 2$. Reducing such an $n$-partite game leads to a biased $k$-partite nonlocal game with higher-dimensional inputs and binary outputs. A particularly symmetric case is studied by Pandit et al.~\cite{Pandit_2022} where for $k=3$, all players have the same number of high-dimension inputs. However, it remains an open problem to investigate the strength of nonlocality attainable when all players possess different high-dimensional inputs. Moreover, motivated by the hierarchy of genuine multipartite quantum correlations established by Jia et al.~\cite{Hierarchy_GMNL_20} using Svetlichny’s definition of genuine multipartite nonlocality, it is natural to explore whether a corresponding hierarchy of nonlocal quantum correlations can be formulated within the LOCCG framework. Finally, biased non-causal games have been developed by Bhattacharya and Banik~\cite{Bhattacharya17} based on the notion of non-causal order introduced by Oreshkov, Costa and Brukner~\cite{Oreshkov12}. Can we identify a similar LOCCG model for genuine multipartite non-causal bounds, which reduce to bipartite biased non-causal games and do these bounds match  bounds for the nonlocal case?

\bibliographystyle{iopart-num}

\appendix
\section{Semidefinite Programming}\label{SDP}

We prove the optimality of the quantum bounds by using an SDP due to Tsirelson's vectorization~\cite{Tsirel80}. We summarize the formulation here and refer the reader to~\cite{Boyd, Wehner08} for details. The primal and equaivalent dual problems may be stated as
\begin{align*}
\label{eq:sdp1}
    \textbf{Primal} & \hphantom{\iff}\textbf{Dual}\\
\max\limits_{G} \frac{1}{2^{n+1}}\Tr(GW) & \iff \min\limits_{\lambda} \lambda \cdot b \\
\text{subject to } G \succcurlyeq 0 &\hphantom{\iff }  \text{ subject to } K \succcurlyeq 0, \\
g_{ii}=1 &
\end{align*}
 where $G$ is the Gram matrix of vectors $\{u_1,\ldots u_{n-k+1}\}$ for Alice and vectors $\{ v_1, \ldots,v_{k+1}\}$ for Bob. Each vector is in $\mathbb{R}^{2^n}$ for a general grouping $(n-k,k)$. We have $G=B^TB$ where 
 \begin{equation*}
 B=\begin{bmatrix}
     u_1&
     u_2&
     \cdots&
     u_{n-k+1}&
     v_1&
     v_2&
     \cdots&
     v_{k+1}
 \end{bmatrix}.    
 \end{equation*}
$G$ is positive semidefinite by construction and having its diagonal entries equal one guarantees that vectors are of unit length. The matrix $W$ is symmetric, with entries that correspond to the Hamming weight of input strings. The vector $b\in \mathbb{R}^{(n+2)}$ in the dual is the all one vector, $\lambda \in \mathbb{R}^{(n+2)}$ is the dual variable and the matrix $K$ is defined as $K=2\text{diag}(\lambda)-W.$

\section{Constructing Observables from the SDP}\label{OBS}
In order to extract the observables from the primal SDP solution matrix $G$, the literature provides algorithms that use Weyl–Brauer operators~\cite{Avis08, Wehner08, Watrous}. We present here a customized application of the alogrithm for our SDP. Let $\ket{\psi}$ be the entangled state and let $D_i$ and $C_j$ represent the observables corresponding to the vectors $u_i$ and $v_j$ such that
 \begin{equation*}
 u_i=D_i  \otimes \mathbb{I}^k \ket{\psi} \textrm{ and } v_j= \mathbb{I}^{n-k} \otimes C_j \ket{\psi}.   
 \end{equation*}
This allows us to equate the expected values with the inner products, giving us $\bra{\psi} D_i \otimes C_j \ket{\psi}= \langle u_i | v_j \rangle$. Our algorithm works even if the dimensions of the observables $D_i$ and $C_j$ are not equal. Note that the solution matrix $G$ has block matrix form given by
\begin{equation*}
 G=\begin{bmatrix}
    P & N\\
    N^T & Q
\end{bmatrix},
\end{equation*}    
where each entry of the $(n-k+1) \times (k+1)$ matrix $N =[n_{ij}]=u_i \cdot v_j$ represents an expected value. The matrices $P$ and $Q$ have dimension $(n-k+1) \times (n-k+1)$ and $(k+1) \times (k+1)$ respectively, while $N^T$ is the transpose of matrix $N$. Let $n'_{ij} = \arccos n_{ij}$ be the angle between vector $u_i$ and $v_j$. Similarly, let $p'_{ij}$ and $q'_{ij}$ be the angle between vectors $u_i, u_j$ and $v_i, v_j$ respectively.

The observables of the optimal protocol are now defined as follows
\begin{align*}\label{obsD}
        D_i&= \cos\theta_i X - \sin\theta_i Z \textrm{ and} \\
        C_j&= \cos\phi_j X - \sin\phi_j Z,
\end{align*}
where $1 \leq i \leq n-k+1$ and $1 \leq j \leq k+1$. Let $\theta_i=p'_{1i}$ be the angle for the observables $D_i$. We fix the angle  $\theta_1=0$ for $D_1$. Similarly, let $\phi_j$ be the angle for the observables $C_j$ such that $\phi_j=n'_{1j}$. We can verify that the expected values in the matrix $N$ correspond to the relation
\begin{equation*}
n_{ij}=\bra{\psi} D_i \otimes C_j \ket{\psi}= \cos(\theta_i- \phi_j).    
\end{equation*}
Building on this relationship, the entries of the dual solution vector $\lambda \in \mathbb{R}^{n+2}$ can be expressed as
 \begin{align*}
     \lambda_i &= \frac{\binom{n-k}{i-1}}{2}\sum_{j=1}^{k+1} f_{ij}\binom{k}{j-1} \cos(\theta_i - \phi_j) \textrm{ and} \\
     \lambda_{j+n-k+2} &= \frac{\binom{k}{j-1}}{2}\sum_{i=1}^{n-k+1} f_{ij}\binom{n-k}{i-1} \cos(\theta_i - \phi_j),
 \end{align*}
where $f_{ij}= (-1)^{f_t(i-1,j-1)}$.

\section{An Example Reduction of LOCCG to Biased Game for MAJORITY}\label{eg:maj}

\begin{figure}
    \centering
    \includegraphics[width = 0.4\linewidth]{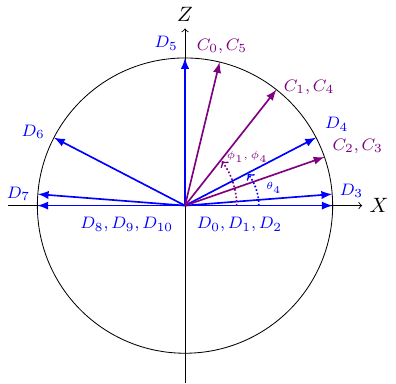}
    \caption{Angles $\theta_i$ and $\phi_j's$ corresponding to the observables $D_i$ and $C_j$ are respectively mapped to the unit circle. The horizontal and vertical axis are aligned with the $X$ and $Z$ operators respectively.}
    \label{unit_circle}
\end{figure}

We begin with the LOCCG model for $MAJORITY$. Let the number of players $n=15$, with the grouping $(10,5)$. Player $i$ and $j$ in groups $A$ and $B$ receive an input bit $x_i, y_j \in \{0,1\}$ and output a bit $a_i, b_j \in \{0,1\}$ respectively, with $1 \leq i \leq 10$ and $1 \leq j \leq 5.$ The winning criteria of the $MAJORITY$ game is 
$$f(x_1, \ldots, x_{10}, y_1, \ldots, y_5)= \begin{cases}
    1 & |x| + |y| \geq \frac{15}{2} \\
    0 & \text{otherwise}
\end{cases}$$
where the threshold $t=\frac{n}{2}$, $|x|=\sum_{i=1}^{10} x_i$ and $|y|=\sum_{j=1}^{5} y_j.$ The players in each group communicate their input to selected players Alice and Bob. All players other than Alice and Bob always output $0$. Alice and Bob obtain their outputs $a$ and $b$ by measuring their shared entangled state. The LOCCG probability distribution for this strategy has the form
$$p(ab a_1 \ldots a_{9} b_1 \ldots b_4 )=\prod_{i=1}^9 p(a_i|x_i) \prod_{j=1}^4 p(b_j|y_j) p(ab|x_1 \ldots x_{10} y_1 \ldots y_5)$$
where
$$p(a_i|x_i)=\begin{cases}
    1 & \text{if } a_i=0\\
    0 & \text{otherwise}
\end{cases} \quad \textrm{and} \quad p(b_j|y_j)=\begin{cases}
    1 & \text{if } b_j=0\\
    0 & \text{otherwise.}
\end{cases}$$
Let $D_{|x|}$ and $C_{|y|}$ be Alice and Bob's observables respectively. 
The optimal choice for these observables, obtained by the SDP in Appendices~\ref{SDP} and~\ref{OBS}, is given by 
\begin{align*}
    D_0&=D_1=D_2=-D_8=-D_9=-D_{10}=X,  && C_0=\frac{1}{2}\sqrt{\frac{11}{50}}X-\frac{3}{2}\sqrt{\frac{21}{50}}Z,\\
    D_3&=\frac{1}{10}\sqrt{\frac{497}{5}}X-\frac{1}{10}\sqrt{\frac{3}{5}}Z, && C_1=\frac{1}{20}\sqrt{\frac{767}{5}}X-\frac{3}{20}\sqrt{\frac{137}{5}}Z,\\
    D_4&=\frac{1}{5}\sqrt{\frac{197}{10}}X-\frac{1}{5}\sqrt{\frac{53}{10}}Z, && C_2=\frac{7}{50}\sqrt{\frac{91}{2}}X-\frac{1}{50}\sqrt{\frac{541}{2}}Z,\\
    D_5&=-Z, && 
    C_3=\frac{7}{50}\sqrt{\frac{91}{2}}X-\frac{1}{50}\sqrt{\frac{541}{2}}Z,\\
    D_6&=-\frac{1}{5}\sqrt{\frac{197}{10}}X-\frac{1}{5}\sqrt{\frac{53}{10}} Z, &&
    C_4=\frac{1}{20}\sqrt{\frac{767}{5}}-\frac{3}{20}\sqrt{\frac{137}{5}}Z,\\
    D_7&=-\frac{1}{10}\sqrt{\frac{497}{5}}X-\frac{1}{10}\sqrt{\frac{3}{5}}Z, &&
    C_5=\frac{1}{2}\sqrt{\frac{11}{50}}X-\frac{3}{2}\sqrt{\frac{21}{50}}Z,
\end{align*}
where $X$ and $Z$ are the Pauli operators. The relationship between these observables is visualized in Figure~\ref{unit_circle}. 
The LOCCG quantum value (\ref{QVal}) of the game is computed as
$$V_Q=\frac{1}{2^{15}} \langle \psi | M | \psi \rangle$$
where
{\small
\begin{align*}\label{BellOp}
    M&= D_0 \otimes C_0  + 5 D_0 \otimes C_1  + 10  D_0 \otimes C_2  + 10  D_0 \otimes C_3  + 5  D_0 \otimes C_4 +  D_0 \otimes C_5  \\
    &+10  D_1 \otimes C_0  + 50 D_1 \otimes C_1  + 100  D_1 \otimes C_2  + 100 D_1 \otimes C_3 + 50 D_1 \otimes C_4 + 10 D_1 \otimes C_5 \\
    &+45 D_2 \otimes C_0 + 225 D_2 \otimes C_1  + 450  D_2 \otimes C_2 + 450 D_2 \otimes C_3 + 225 D_2 \otimes C_4 + 45 D_2 \otimes C_5\\
    &+120 D_3 \otimes C_0 + 600 D_3 \otimes C_1 + 1200  D_3 \otimes C_2 + 1200  D_3 \otimes C_3  + 600  D_3 \otimes C_4 \\
    & - 120 D_3 \otimes C_5+210  D_4 \otimes C_0  + 1050 D_4 \otimes C_1 + 2100  D_4 \otimes C_2  + 2100  D_4 \otimes C_3   \\
    &- 1050  D_4 \otimes C_4 - 210 D_4 \otimes C_5+252  D_5 \otimes C_0  + 1260 D_5 \otimes C_1 + 2520  D_5 \otimes C_2   \\
    &- 2520  D_5 \otimes C_3  - 1260  D_5 \otimes C_4 - 252 D_5 \otimes C_5+210  D_6 \otimes C_0  + 1050 D_6 \otimes C_1  \\
    &- 2100  D_6 \otimes C_2  - 2100  D_6 \otimes C_3  - 1050  D_6 \otimes C_4 - 210 D_6 \otimes C_5+120 D_7 \otimes C_0 \\
    & - 600 D_7 \otimes C_1 - 1200  D_7 \otimes C_2 - 1200  D_7 \otimes C_3  - 600  D_7 \otimes C_4 - 120 D_7 \otimes C_5  \\
    &-45 D_8 \otimes C_0 - 225 D_8 \otimes C_1  - 450  D_8 \otimes C_2 - 450 D_8 \otimes C_3 - 225 D_8 \otimes C_4 - 45 D_8 \otimes C_5\\
    &-10  D_9 \otimes C_0  - 50 D_9 \otimes C_1  - 100  D_9 \otimes C_2  - 100 D_9 \otimes C_3 - 50 D_9 \otimes C_4 - 10 D_9 \otimes C_5 \\
    &-D_{10} \otimes C_0  - 5 D_{10} \otimes C_1  - 10  D_{10} \otimes C_2  - 10  D_{10} \otimes C_3  - 5  D_{10} \otimes C_4 -  D_{10} \otimes C_5.
\end{align*}
}
Plugging in the observables, the optimal quantum value of the game is $V_Q^*=0.670201$. This is strictly higher than the optimal classical value of the game provided in the table in Figure~\ref{tab1}. 

Note that $|x| \in \{0,1,2\}$ are the Hamming weights for which Alice knows with certainty that $|x|+|y| < 7$. Similarly, for $|x| \in \{8,9,10\}$, the $MAJORITY$ threshold is achieved,~i.e.,~ $|x|+|y| \geq 8$. In both cases, Alice has enough information to determine whether the input satisfies $MAJORITY$. This also shows up in the choice of optimal observables in the equivalence
\begin{equation*}
D_0=D_1=D_2=-D_8=-D_9=-D_{10}.
\end{equation*}
We replace $D_1$ and $D_2$ with ${D}_0$ in the Bell Operator $M$ and $D_8, D_9$ and $D_{10}$ with $-{D}_0$. This allows us to rewrite the operator $M$ as
{\small
\begin{align*}\label{BellOp:eg}
    M&= 112 D_0 \otimes C_0  + 560 D_0 \otimes C_1  + 1120  D_0 \otimes C_2  + 1120  D_0 \otimes C_3  + 560  D_0 \otimes C_4   \\
    &+  112D_0 \otimes C_5+120 D_3 \otimes C_0 + 600 D_3 \otimes C_1 + 1200  D_3 \otimes C_2 + 1200  D_3 \otimes C_3    \\
    &+ 600  D_3 \otimes C_4- 120 D_3 \otimes C_5+210  D_4 \otimes C_0  + 1050 D_4 \otimes C_1 + 2100  D_4 \otimes C_2    \\
    &+ 2100  D_4 \otimes C_3  - 1050  D_4 \otimes C_4- 210 D_4 \otimes C_5+252  D_5 \otimes C_0  + 1260 D_5 \otimes C_1  \\
    &+ 2520  D_5 \otimes C_2  - 2520  D_5 \otimes C_3  - 1260  D_5 \otimes C_4 - 252 D_5 \otimes C_5+210  D_6 \otimes C_0\\
    &  + 1050 D_6 \otimes C_1 - 2100  D_6 \otimes C_2  - 2100  D_6 \otimes C_3  - 1050  D_6 \otimes C_4 - 210 D_6 \otimes C_5 \\
    &+120 D_7 \otimes C_0 - 600 D_7 \otimes C_1 - 1200  D_7 \otimes C_2 - 1200  D_7 \otimes C_3  - 600  D_7 \otimes C_4 \\
    &- 120 D_7 \otimes C_5.
\end{align*}
}

\begin{figure}
    \centering
    \includegraphics[width = 0.2\linewidth]{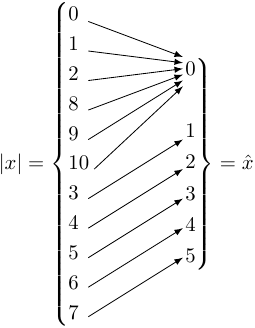}
    \caption{ The mapping of LOCCG input strings $x \in \{0,1\}^{10}$ to integer inputs $\hat{x}$ for biased $MAJORITY$.}
    \label{eg_(10,5)}
\end{figure}
Based on these observations, we construct the biased game between Alice and Bob by redefining the inputs to Alice and Bob. Using Equations~\ref{hatinput} and~\ref{beta}, these are now given by  $0 \leq \hat{x} \leq 5$ and $0 \leq \hat{y} \leq 5$ where $\hat{y}=|y|$. The original inputs of length $|x| \in \{0,1,2,8,9,10\}$ are mapped to $\hat{x}=0$ and the remaining inputs of length $|x| \in \{3,4,5,6,7\}$ are mapped to $\hat{x} \in \{1,2,3,4,5\}$ as shown in Figure \ref{eg_(10,5)}. The new winning condition based on Equation~\ref{Bcriteria} is 
\begin{align*}
    g(\hat{x},\hat{y})=\begin{cases}
        1 & \text{if } \hat{x} + \hat{y} > 5, \text{ and }\\
        0 & \text{otherwise.}
    \end{cases}
\end{align*}
The multipartite $MAJORITY$ game for $(10,5)$ is now reduced to the bipartite biased $MAJORITY$ game with integer inputs. The optimal protocol of the biased game is given by
\begin{align*}
    \hat{D}_0&=X,  && \hat{C}_0=\frac{1}{2}\sqrt{\frac{11}{50}}X-\frac{3}{2}\sqrt{\frac{21}{50}}Z,\\
    \hat{D}_1&=\frac{1}{10}\sqrt{\frac{497}{5}}X-\frac{1}{10}\sqrt{\frac{3}{5}}Z, && \hat{C}_1=\frac{1}{20}\sqrt{\frac{767}{5}}X-\frac{3}{20}\sqrt{\frac{137}{5}}Z,\\
    \hat{D}_2&=\frac{1}{5}\sqrt{\frac{197}{10}}X-\frac{1}{5}\sqrt{\frac{53}{10}}Z, && \hat{C}_2=\frac{7}{50}\sqrt{\frac{91}{2}}X-\frac{1}{50}\sqrt{\frac{541}{2}}Z,\\
    \hat{D}_3&=-Z, && 
    \hat{C}_3=\frac{7}{50}\sqrt{\frac{91}{2}}X-\frac{1}{50}\sqrt{\frac{541}{2}}Z,\\
    \hat{D}_4&=-\frac{1}{5}\sqrt{\frac{197}{10}}X-\frac{1}{5}\sqrt{\frac{53}{10}} Z, &&
    \hat{C}_4=\frac{1}{20}\sqrt{\frac{767}{5}}-\frac{3}{20}\sqrt{\frac{137}{5}}Z,\\
    \hat{D}_5&=-\frac{1}{10}\sqrt{\frac{497}{5}}X-\frac{1}{10}\sqrt{\frac{3}{5}}Z, &&
    \hat{C}_5=\frac{1}{2}\sqrt{\frac{11}{50}}X-\frac{3}{2}\sqrt{\frac{21}{50}}Z,
\end{align*}
where the observables $\hat{D}_{\hat{x}}$ and $\hat{C}_{\hat{y}}$ of Alice and Bob have been relabeled according to Equation~\ref{hatinput}. We can verify that the optimal quantum value of the biased game (given by Equation~\ref{BiasedQVal}) is equal to the multipartite $MAJORITY$ game in LOCCG model with grouping $(10,5)$, i.e.~$V_Q^*=0.670201.$

\begin{table*}[t]  
     \centering
     \resizebox{\hsize}{!}{
     \begin{tabular}{cccccc}
     \hline
     \hline
       $n-2$ & Primal Solution Optimal Strategy & Dual Solution Vector $\lambda$ & $V^*_{Q}$ &$V^*_{C}$ & ${V^*_{Q}}/{V^*_{C}}$   \\
       \hline
         $2$ & 
     $\begin{matrix}
      D_0 =& X\\
     D_1 =&  -\frac{1}{11}\sqrt{\frac{599}{500}}{X}-\frac{1}{11}\sqrt{\frac{59901}{500}}{Z}\\
     D_2 =& -\frac{3}{55}\sqrt{\frac{311}{2}}{X}-\frac{1}{55}\sqrt{\frac{3251}{2}}{Z}\\
     C_0=&\frac{1}{4}\sqrt{\frac{93}{10}}X -\frac{1}{4}\sqrt{\frac{67}{10}}{Z}\\
     C_1=&\frac{1}{40}\sqrt{\frac{2581}{5}}{X} + \frac{1}{40}\sqrt{\frac{5419}{5}}{Z}\\
     C_2=&-\frac{1}{110}\sqrt{\frac{237}{10}} {X} + \frac{1}{110}\sqrt{\frac{120763}{10}}  Z\\
     \end{matrix}$
     & $\frac{1}{16}\begin{pmatrix}
                \frac{1}{8}\sqrt{\frac{93}{10}}+\frac{1}{40}\sqrt{\frac{2581}{5}}+\frac{1}{220}\sqrt{\frac{237}{10}}\\
                \frac{1}{11}\sqrt{\frac{599}{500}}\{\frac{1}{20}\sqrt{\frac{2581}{5}}-\frac{1}{4}\sqrt{\frac{93}{10}}-\frac{1}{110}\sqrt{\frac{237}{10}}\}+
                \frac{1}{11}\sqrt{\frac{59901}{500}}\{\frac{1}{20}\sqrt{\frac{5419}{5}}+\frac{1}{4}\sqrt{\frac{67}{10}}+\frac{1}{110}\sqrt{\frac{120763}{10}}\}
                \\
                \frac{3}{22}\sqrt{\frac{311}{2}}\{\frac{1}{100}\sqrt{\frac{2581}{5}}-\frac{1}{550}\sqrt{\frac{237}{10}}+\frac{1}{20}\sqrt{\frac{93}{10}}\}+
                \frac{1}{22}\sqrt{\frac{3251}{2}}\{\frac{1}{100}\sqrt{\frac{5419}{5}}+\frac{1}{550}\sqrt{\frac{120763}{10}}-\frac{1}{20}\sqrt{\frac{67}{10}}\}
                \\
                \frac{1}{8}\sqrt{\frac{93}{10}}+\frac{1}{44}\sqrt{\frac{59901}{500}}\sqrt{\frac{67}{10}}-\frac{1}{44}\sqrt{\frac{599}{500}}\sqrt{\frac{93}{10}}+\frac{3}{440}\sqrt{\frac{311}{2}}\sqrt{\frac{93}{10}}-\frac{1}{440}\sqrt{\frac{3251}{2}}\sqrt{\frac{67}{10}}
                \\
                \frac{1}{40}\sqrt{\frac{2581}{5}}+\frac{1}{220}\sqrt{\frac{599}{500}}\sqrt{\frac{2581}{5}}+\frac{1}{220}\sqrt{\frac{59901}{500}}\sqrt{\frac{5419}{5}}
                +\frac{3}{2200}\sqrt{\frac{311}{2}}\sqrt{\frac{2581}{5}}+\frac{1}{2200}\sqrt{\frac{3251}{2}}\sqrt{\frac{5419}{5}}
                \\
                \frac{1}{110}\sqrt{\frac{237}{10}}+\frac{1}{1210}\sqrt{\frac{59901}{500}}\sqrt{\frac{120763}{10}}-\frac{1}{1210}\sqrt{\frac{599}{500}}\sqrt{\frac{237}{10}}+\frac{1}{12100}\sqrt{\frac{3251}{2}}\sqrt{\frac{120763}{10}}-\frac{3}{12100}\sqrt{\frac{311}{2}}\sqrt{\frac{237}{10}}
     \end{pmatrix}$
      & $0.7054$
      &$0.625$
      &$1.1286$\\
     \hline
     $3$ & $\begin{matrix}
     D_0 =& X\\
     D_1 =& \frac{\sqrt{5}}{3} X - \frac{2}{3} Z \\
     D_2=& -\frac{\sqrt{5}}{3} X- \frac{2}{3}Z\\
     C_0 =& \frac{1}{\sqrt{5}} X - \frac{2}{\sqrt{5}}Z\\
     C_1 =& X \\
     C_2=& \frac{1}{\sqrt{5}}{X} + \frac{2}{\sqrt{5}}{Z}\\
     \end{matrix}$ 
     &$\frac{1}{32\sqrt{5}}\begin{pmatrix}
         2+2\sqrt(5)\\
         9\\
         9\\
         5\\
         10+2\sqrt{5}\\
         5
     \end{pmatrix}$
      & 
      $0.684$
      &$0.625$
      &$1.0944$\\
     \hline
     $4$& 
       $\begin{matrix}
     D_0 =& X\\
     D_1 =& \frac{1}{30}\sqrt{\frac{3103}{5}} {X} -\frac{1}{30}\sqrt{\frac{1397}{5}}Z \\
     D_2 =& -\frac{8}{15}\sqrt{\frac{23}{10}} {X}-\frac{1}{15} \sqrt{\frac{389}{5}} {Z} \\
     C_0 =& \frac{1}{50}\sqrt{\frac{4699}{5}}X - \frac{1}{50}\sqrt{\frac{7801}{5}}Z\\
     C_1 =& \frac{1}{6}\sqrt{\frac{357}{10}}X + \frac{1}{6}\sqrt{\frac{3}{10}}Z \\
     C_2=& \frac{\sqrt{142}}{15}X + \frac{\sqrt{83}}{15}Z\\
     \end{matrix}$ 
     &$\frac{1}{2^6}
     \begin{pmatrix}
        \sqrt{\frac{357}{10}} + \frac{\sqrt{142}}{5} + \frac{3}{50} \sqrt{\frac{4699}{5}}\\   
   \frac{\sqrt{2215542}-\sqrt{8382}}{450} + \frac{\sqrt{579755} -  \sqrt{2203130}}{1125} + \frac{\sqrt{10897997} + \sqrt{14580997})}{3750}\\
     \frac{\sqrt{2334}}{150} + \frac{8\sqrt{8165}+ \sqrt{161435}}{375}  + \frac{4 \sqrt{8211}}{5}  - \frac{2 \sqrt{216154}}{625} + \frac{\sqrt{3034589}}{1250}   \\
    \frac{3 \sqrt{23495}}{250} - \frac{2\sqrt{216154}}{625}  + \frac{\sqrt{3034589}}{1250} + \frac{\sqrt{  10897997} + \sqrt{14580997}}{3750}  \\ 
    \frac{\sqrt{2334}}{150} + \frac{\sqrt{3570}}{10} + \frac{4\sqrt{8211}}{75}  -\frac{(\sqrt{8382} - \sqrt{2215542})}{450}\\
   \frac{\sqrt{142}}{5} + \frac{8\sqrt{8165}}{375}  + \frac{\sqrt{161435}}{375}  + \frac{\sqrt{579755} - \sqrt{2203130}}{1125}
     \end{pmatrix}$
      &$ 0.7067$
      &$0.6875$
      &$1.0279$\\
     \hline
    $5$   & 
     $\begin{matrix}
     D_0 =& X\\
     D_1 =& \frac{\sqrt{17}}{5} X - \frac{2\sqrt{2}}{5} Z &  \\
     D_2=& -\frac{\sqrt{17}}{5} {X}- \frac{2\sqrt{2}}{5}{Z}\\
     C_0 =& \frac{1}{5}\sqrt{\frac{397}{30}} X - \frac{1}{5}\sqrt{\frac{353}{30}} Z\\
     C_1 =& X \\
     C_2=& \frac{1}{5}\sqrt{\frac{397}{30}} X + \frac{1}{5}\sqrt{\frac{353}{30}} Z\\
     \end{matrix} $
     & $\frac{1}{2^7}\begin{pmatrix}
     \frac{12}{5}\sqrt{\frac{397}{30}}+12 \\ \frac{4}{5}\sqrt{\frac{353}{15}}+2\sqrt{17} \\ \frac{4}{5}\sqrt{\frac{353}{15}}+2\sqrt{17} \\ \frac{6}{5}\sqrt{\frac{397}{30}}+\frac{4}{5}\sqrt{\frac{353}{15}} \\ 4\sqrt{17}+12 \\ \frac{6}{5}\sqrt{\frac{397}{30}}+\frac{4}{5}\sqrt{\frac{353}{15}}
     \end{pmatrix}$  
     &  
          $0.7028$
          &$0.6875$
     & $1.0223$\\
     \hline
     $6$ &  $\begin{matrix}
     D_0 =& X \\
     D_1 =& \frac{1}{20}\sqrt{\frac{3427}{10}}X - \frac{1}{20}\sqrt{\frac{573}{10}}Z  \\
     D_2 =&  -\frac{1}{20}\sqrt{\frac{3217}{10}}X-\frac{3}{20}\sqrt{\frac{87}{10}}Z\\
     C_0 =& \frac{1}{20}\sqrt{\frac{1489}{5}}X-\frac{1}{20}\sqrt{\frac{511}{5}}Z\\
     C_1 =& \frac{\sqrt{399}}{20}X+\frac{1}{20}Z\\
     C_2=& \frac{3}{10}\sqrt{\frac{93}{10}}X+\frac{1}{10}\sqrt{\frac{163}{10}}\\
     \end{matrix}$
     &$\frac{1}{2^8}\begin{pmatrix}
     \frac{87}{20}\sqrt{\frac{93}{10}}+\frac{29}{40}\sqrt{\frac{1489}{5}}+\frac{29}{20}\sqrt{399}\\
     \frac{3\sqrt{93399}-3\sqrt{5730}-9\sqrt{318711}}{800}   + \frac{3\sqrt{585606} +3 \sqrt{10205606} + 6 \sqrt{13673730}}{1600}\\
   \frac{3 \sqrt{870} + 3 \sqrt{14181} + 3 \sqrt{299181}}{200} + \frac{3 \sqrt{88914}}{400} - \frac{\sqrt{9580226}}{400} + \frac{ \sqrt{12835830}}{200}\\
     \frac{29\sqrt{7445}}{200} + \frac{3\sqrt{88914}}{400} + \frac{3 \sqrt{585606}+ 3 \sqrt{10205606}}{1600} - \frac{\sqrt{9580226}}{400} \\
    \frac{29\sqrt{399}}{20} + \frac{3(\sqrt{870}-\sqrt{5730}+\sqrt{13673730})}{200} +  \frac{\sqrt{12835830}}{200}  \\
   \frac{87\sqrt{930}}{200}  + \frac{3(\sqrt{14181}+\sqrt{299181})}{200} + \frac{3\sqrt{93399}}{800}  -\frac{9\sqrt{318711}}{800}
    \end{pmatrix}$ 
     &$0.7296$ 
     &$0.7266$
     & $1.0041$\\
     \hline 
     $7$ &  $\begin{matrix}
    D_0 =& X\\
     D_1 =& \frac{1}{60}\sqrt{\frac{6071}{2}}{X} - \frac{1}{60}\sqrt{\frac{1129}{2}}{Z} \\
     D_2=& -\frac{1}{60}\sqrt{\frac{6071}{2}}{X}- \frac{1}{60}\sqrt{\frac{1129}{2}}{Z}\\
     C_0 =& \frac{1}{4}\sqrt{\frac{1989}{161}}X - \frac{1}{4}\sqrt{\frac{587}{161}}Z\\
     C_1 =& X\\
     C_2=& \frac{1}{4}\sqrt{\frac{1989}{161}}X + \frac{1}{4}\sqrt{\frac{587}{161}}Z\\
     \end{matrix}$ 
     &$\frac{1}{2^9} \begin{pmatrix}
     58+\frac{87}{2}\sqrt{\frac{221}{161}}\\
    \frac{7}{48}\sqrt{\frac{1129}{2}}\sqrt{\frac{587}{161}}+\frac{7}{12}\sqrt{\frac{6071}{2}}\\
      \frac{7}{48}\sqrt{\frac{1129}{2}}\sqrt{\frac{587}{161}}+\frac{7}{12}\sqrt{\frac{6071}{2}}\\
      \frac{87}{4} \sqrt{\frac{221}{161}} + \frac{1}{48}\sqrt{\frac{4639061}{46}}\\
      58+\frac{7}{6}\sqrt{\frac{6071}{2}}\\
      \frac{87}{4} \sqrt{\frac{221}{161}} + \frac{1}{48}\sqrt{\frac{4639061}{46}}
     \end{pmatrix}$
      & 
 $0.7288$
      &$0.7266$
      & $1.003$\\
     \hline
     $8$ &
    $\begin{matrix}
         D_0 =&{X}\\
        D_1 =&{X}\\
        D_2 =&-{X}\\
        C_0=&C_1=C_2=X\\
    \end{matrix}$ 
    &$\frac{1}{2^{10}}\begin{pmatrix}
    260\\
56\\
70\\
58\\
256\\
72
    \end{pmatrix}$ 
    &$0.7539 $
    &$0.7539$
    &$1$\\
    \hline \hline
     \end{tabular}
     }
     \caption{\label{tab3}The optimal quantum strategies and matching dual solutions for $MAJORITY$ with  $n \in \{4, \ldots ,10 \}$ and $k=2$,~i.e.,$(n-2,2)$.}
\end{table*}

\end{document}